\documentclass[10pt,journal]{IEEEtran}

\usepackage{cite}
\usepackage[justification=centering]{caption}

\usepackage[pdftex]{graphicx}

\usepackage{url}
\usepackage{algorithm}  
\usepackage{algorithmicx}  
\usepackage{algpseudocode}
\usepackage[namelimits]{amsmath} 
\usepackage{amssymb}             
\usepackage{amsfonts}            
\usepackage{mathrsfs} 
\usepackage{amssymb,amsmath,epsfig}
\usepackage{textcomp}
\usepackage{amsfonts,amsthm,amsopn}
\usepackage{hyperref}
\usepackage{adjustbox}
\usepackage{tikz}

\newcommand{\vct}[1]{\boldsymbol{\mathbf{#1}}}
\usepackage{enumerate}   
\usepackage[normalem]{ulem}
\usepackage{soul}
\usepackage{url}
\usepackage[capitalise]{cleveref}
\usepackage{graphicx,caption,lipsum}
\usepackage{booktabs,multirow}
\usepackage{placeins}
\usepackage{algpseudocode,algorithm}
\usepackage{color}

\usepackage{enumerate}   
\usepackage[normalem]{ulem}

\usepackage{subfigure}
\usepackage{array}

\usepackage{booktabs}
 \usepackage{siunitx}
\usepackage{tikz,xcolor,hyperref}
\definecolor{lime}{HTML}{A6CE39}
\DeclareRobustCommand{\orcidicon}{%
	\begin{tikzpicture}
	\draw[lime, fill=lime] (0,0) 
	circle [radius=0.16] 
	node[white] {{\fontfamily{qag}\selectfont \tiny ID}};	\draw[white, fill=white] (-0.0625,0.095) 
	circle [radius=0.007];	\end{tikzpicture}
	\hspace{-2mm}}
\foreach \x in {A, ..., Z}{%
	\expandafter\xdef\csname orcid\x\endcsname{\noexpand\href{https://orcid.org/\csname orcidauthor\x\endcsname}{\noexpand\orcidicon}}
	}

\usepackage{footnote}
\makesavenoteenv{tabular}
\usepackage{caption} 
\usepackage{float}

\begin{document}
%
\title{Multi-source Domain Adaptation for Text-independent Forensic  Speaker Recognition}
%
%
%
\author{Zhenyu~Wang, 
        and~John~H.~L.~Hansen\orcidA{}, \textit{Fellow}, \emph{IEEE}
\thanks{Manuscript received April 1, 2021; revised July 27, 2021 and November 8, 2021; accepted November 15, 2021. Date of publication November 26, 2021; date of current version December 20, 2021. This work was supported by the University of Texas at Dallas from the Distinguished University Chair in Telecommunications Engineering held by John H. L. Hansen. The associate editor coordinating the review of this manuscript and approving it for publication was Dr. Alberto Abad. (Corresponding author: John H. L. Hansen.)}
\thanks{The authors are with Center for Robust Speech Systems, Erik Jonsson School of Engineering University of Texas at Dallas, Richardson, TX 75080, USA (e-mail: zhenyu.wang@utdallas.edu; john.hansen@utdallas.edu). 
}
\thanks{Digital Object Identifier 10.1109/TASLP.2021.3130975}}

\markboth{IEEE/ACM TRANSACTIONS ON AUDIO, SPEECH, AND LANGUAGE PROCESSING}
{Shell \MakeLowercase{\textit{et al.}}: Bare Demo of IEEEtran.cls for IEEE Journals}
%



\maketitle

\begin{abstract}
Adapting speaker recognition systems to new environments is a widely-used technique to improve a well-performing model learned from large-scale data towards a task-specific small-scale data scenarios. However, previous studies focus on single domain adaptation, which neglects a more practical scenario where training data are collected from multiple acoustic domains needed in forensic scenarios. Audio analysis for forensic speaker recognition offers unique challenges in model training with multi-domain training data due to location/scenario uncertainty and diversity mismatch between reference and naturalistic field recordings. It is also difficult to directly employ small-scale domain-specific data to train complex neural network architectures due to domain mismatch and performance loss. Fine-tuning is a commonly-used method for adaptation in order to retrain the model with weights initialized from a well-trained model. Alternatively, in this study, three novel adaptation methods based on domain adversarial training, discrepancy minimization, and moment-matching approaches are proposed to further promote adaptation performance across multiple acoustic domains. A comprehensive set of experiments are conducted to demonstrate that: 1) diverse acoustic environments do impact speaker recognition performance, which could advance research in audio forensics, 2) domain adversarial training learns the discriminative features which are also invariant to shifts between domains, 3) discrepancy-minimizing adaptation achieves effective performance simultaneously across multiple acoustic domains, and 4) moment-matching adaptation along with dynamic distribution alignment also significantly promotes speaker recognition performance on each domain, especially for the LENA-field domain with noise compared to all other systems. Advancements shown here in adaptation therefore helper ensure more consistent performance for field operational data in audio forensics.
\end{abstract}


%
\maketitle

\section{Introduction}
%
%
%
%
\IEEEPARstart{I}{n} general, no two speakers are identical, differing in anatomy, physiology, and acoustically from a speech production viewpoint. Considering human speech as a discriminative biometric, speaker recognition serves as an important tool in law enforcement, national security, and forensics in general. The need for forensic speaker recognition arises when an individual contributes his/her voice as evidence, including telephone recordings, wiretaps, audio surveillance, or informant recordings \cite{hansen2015speaker}. The use of technology for forensic speaker recognition has been considered as early as 1926 based on speech waveform analysis \cite{wigmore1926new}. It was popularized much later in the 1970s, when it came to be known incorrectly as the "voiceprint" \cite{kersta1962voiceprint}. Approaches to forensic speaker recognition include spectrographic, auditory, acoustic-phonetic, and automatic. Forensic speaker recognition is commonly performed fully or partially by human expert phoneticians who generally have backgrounds in linguistics and statistics. Full or assisted automatic approaches are also considered as an efficient tool for forensic speaker recognition to aid the forensic examiner in quantifying the strength of evidence\cite{pop2014quality,boe2000forensic,castro2007forensic,bonastre2003person}.

In the forensic context, speaker recognition assists investigators and legal/courtrooms (judge or jury) to identify an unknown speaker suspected of a crime in legal proceedings. In general, for forensic speaker recognition, a likelihood ratio is needed to determine how likely a voice recording was produced by a speaker of known identity (typically a suspect) or not \cite{wigmore1926new,enzinger2016implementation}.

Great progress has been made in speaker recognition in recent decades, thus solidifying automatic speaker recognition as a core tool in the forensic field. Previously, the segment-level vectors that represent speech entitled i-Vectors with probabilistic linear discriminant analysis (PLDA) as a backend have dominated the text-independent speaker recognition research field\cite{dehak2010front}. Additionally, i-Vector variants have been widely used in multiple fields of paralinguistic speech attribute recognition \cite{dehak2011language,li2014speaker,zhang2018text}. With the emergence of large speaker labeled audio datasets and growing computational resources, there is increased interest in applying more effective approaches including x-Vector and other neural network architectures to speaker recognition tasks \cite{snyder2018x,snyder2016deep,li2017deep,lei2014novel,cai2018exploring,kumar2020designing,xia2020self}.

Forensic speech data as potential evidence can be obtained in random naturalistic environments resulting in variable data quality. Additionally, speaker-based uncertainties such as stress, sentiment, vocal effort, and other intrinsic speaker factors introduce unknown mismatch challenges\cite{hansen2015speaker}. Mismatch variability consisting of intrinsic and extrinsic characteristics can degrade performance of speaker recognition. Intrinsic speaker characteristics represent speech traits that are dependent on the speaker vs. extrinsic characteristics that are dependent on audio capture and environmental factors\cite{hansen2015speaker}. Intrinsic properties include the speaker's age, gender, ethnicity, vocal effort, noise-induced Lombard effect \cite{hansen2009analysis}, situational stress, emotional and physical state (e.g., angry, sad, stressed, distracted, etc.). Extrinsic properties include recording equipment conditions such as microphone type and placement, background noise, room reverberation, and other environmental scenario-based issues. Some factors, such as noise and non-target speech, may impact system performance by their mere presence. Variations caused by multi-faceted acoustic scenarios pose major challenges for effective model development to recognize a speaker in mismatched conditions.

To mitigate the impact of condition mismatch or domain shift, domain adaptation is needed to generalize a well-trained model learned for one acoustic domain to encompass new domains with task-specific data. Some theoretical insights can be drawn from the computer vision (CV) field. For example, some domain adaptation methods\cite{long2017deep,tzeng2014deep,long2015learning} employ neural networks with a Maximum Mean Discrepancy (MMD) loss to diminish domain discrepancy; other methods introduce novel model training schema to optimize domain alignments such as moment matching \cite{zellinger2017central}, adversarial domain confusion \cite{tzeng2017adversarial,ganin2015unsupervised,saito2018maximum} and Generative Adversarial Network (GAN)-based alignment \cite{zhu2017unpaired,liu2017unsupervised,hoffman2018cycada}. Such methods have also been adopted by the speaker adaptation community to address tasks focusing on domain mismatch, training techniques, and new architectures. Learning domain-invariant speaker embeddings with GAN have been considered to mitigate the impact of language mismatch between training and evaluation sets\cite{bhattacharya2019generative,xia2019cross}. For example, \cite{lin2020multi,lin2020framework} proposed to adapt speaker recognition systems with MMD at both frame-level and segment-level, which is assumed to be more robust against duration discrepancy. New architectures optimizing with objective-targeting losses have also been investigated to extract robust embeddings \cite{sarfjoo2019domain,vesely2016sequence}. However, the domain adaptation methods noted above mainly align the distributions of representations that may only contain partial information for a single domain, which are not practical in many forensic scenarios where speech samples are typically collected from diverse naturalistic domains with multiple mismatches containing unknown context knowledge, thereby, requiring Multi-source domain adaptation (MSDA). Our previous work \cite{wang2020cross} utilized the discrepancy minimization method for cross-domain adaptation and evaluated system performance within a closed-set using forensic data. Related works have also been applied in the CV field recently, such as novel cross-domain structures based on the formalism of multi-class domain adaptation were proposed. These studies consider the concept of minimizing the domain distance or category shift with measures such as MMD, Moment Distance (MD), and Multi-Class Scoring Disagreement (MCSD) \cite{zhu2019multi,zhang2020unsupervised,xu2018deep,peng2019moment,zhu2019aligning}.

In this study, we develop three multi-source domain adaptation approaches to learn domain-invariant information across naturalistic environments containing extrinsic variations. These variations alter speaker identity traits, whose instantiation variants require new learning objectives which either coincide with or resemble widely-used methods, thus partially underscoring their effectiveness in more practical scenarios. To address the lack of available real naturalistic forensic audio corpora with ground-truth speaker identity, we introduce our CRSS-Forensic dataset for benchmarking state-of-the-art multi-source domain adaptation methods. The dataset includes four subsets: ($i$) Clean (e.g. audio recorded with a close-talk mic and a desk-top mic), ($ii$) Far-field (e.g. audio recorded with distance mics), ($iii$) LENA-booth (e.g. audio recorded with a asynchronous mobile data collection platform called LENA worn by the participant), and ($iv$) LENA-field (e.g. audio recorded in public environments with noise), where three kinds of mismatch such as distance mismatch, channel mismatch, and noise mismatch exist among these subsets. First, an x-Vector system is pre-trained with a large-scale VoxCeleb dataset, followed by fine-tuning the high-level neural network layers to learn speaker information from the CRSS-Forensic corpus. In addition to the pre-trained x-Vector model, we perform a multi-source domain adaptation using two alternative methods. One is based on discrepancy minimization to align the domain-specific distributions with maximum mean discrepancy (MMD). The second employs a moment-matching method to minimize the inter-domain discrepancies and dynamically aligns the moments of embedding distributions with an adversarial training strategy. In terms of our test protocol, we evaluate the pre-trained x-Vector system, fine-tuned system, discrepancy-minimization adaptation system, and moment-matching system with the Phase-1 portion of the CRSS-Forensic dataset under an open-set framework. The main contributions of this study are as follows,
\begin{itemize}
	\item[1)]
we demonstrate the impact of different acoustic environments on speaker recognition system performance.
	\item[2)]
A set of speaker recognition adaptation approaches are proposed to address forensic speaker recognition under diverse acoustic environments.
	\item[3)]
A discussion regarding best practices on how the proposed speaker recognition system could assist the "trier of fact" (i.e., a judge, a panel of judges, or a jury) in making decisions regarding the origin of speech on voice recordings of a speaker whose identity is in question.
\end{itemize}

This paper is organized as follows: Sec. \uppercase\expandafter{\romannumeral2} describes the x-Vector backbone system and fine-tuning details. Sec. \uppercase\expandafter{\romannumeral3} elaborates on our domain adversarial training approach. The description of our proposed adaptation system framework based on discrepancy minimization along with each component are presented in Sec. \uppercase\expandafter{\romannumeral4}. Sec. \uppercase\expandafter{\romannumeral5} elaborates on the proposed system framework based on moment matching and our corresponding adversarial training schema. A brief description of each system's evaluation corpus and configurations are also illustrated, and the dataset description is included in Sec. \uppercase\expandafter{\romannumeral6}. The effectiveness of the proposed methods is demonstrated in Sec. \uppercase\expandafter{\romannumeral7} using a performance comparison across each sub-domain subset of the CRSS-Forensic corpus. Finally, conclusions are summarized in Sec. \uppercase\expandafter{\romannumeral8}.
\maketitle
\section{Backbone System and Fine-tuning}
\vspace{1ex}

Since the x-Vector \cite{snyder2018x} has shown competitive results when trained on large proprietary datasets and is widely accepted as an effective speaker recognition solution, we employ the x-Vector system as the backbone system. Here, we pre-train an x-Vector system with large-scale VoxCeleb data to obtain a preliminary discriminative speaker representation, then fine-tune that model using audio data from the CRSS-Forensic corpus.
\subsection{Pre-trained System}
The time-delay layer $\mathcal F (\rm{N},\rm{D},\rm{K})$ \cite{peddinti2015time} forms the basic component of the x-Vector system which computes fixed-length speaker embeddings from variable-length acoustic segments. At each time-step, activations from the previous layer are computed using a context width of $\rm{K}$, and a dilation of $\rm{D}$. Here, $\rm{N}$ represents the output embedding dimension. The temporal short-term context is processed by this feed-forward time-delay architecture at the frame-level, where the statistics pooling layer is used to aggregate over frame-level representations to compute corresponding mean and standard deviations as the concatenation output. At the segment-level, additional fully-connected layers are used to operate on non-temporal concatenated information followed by a softmax output layer\cite{snyder2017deep}. Ultimately, the goal of this architecture is to generate speaker embeddings over the entire utterance that hopefully will generalize well to unseen speakers within the training set. Therefore, suppose there are $S$ speakers with $M$ training samples, then the training objective is to maximize the probability $P(y_S | {\rm{x}}_{1:T}^{(M)})$ for speaker $S$ given the $T$ input frames ${\rm{x}}_1^{(M)}, {\rm{x}}_2^{(M)}, \dots, {\rm{x}}_T^{(M)}$. The optimization process can then be written as,
\setlength{\arraycolsep}{0.0em}
\begin{equation}
\Theta^{*} = \mathop{\arg\max}_{\Theta} [ \mathbb{E}_{\vct x} [\log p_{\Theta}(y|\vct x)].
\end{equation}
Here, $\Theta = \{\mathbf{W}^l,\mathbf{b}^l\}_{l=1}^L$ denotes the trainable parameters of the $L$-layer neural network, $({\vct x},y)$ represents the frame-level feature of an utterance and its corresponding label.

In this study, we employ the same backend probabilistic linear discriminant analysis (PLDA) \cite{ioffe2006probabilistic} training method for each system. Speaker embeddings are centered and dimensionality reduction is accomplished using Linear Discriminant Analysis (LDA). After LDA, scores for pairs of length-normalized embeddings are generated using the PLDA model and normalized with an adaptive s-norm \cite{sturim2005speaker}. Next, a PLDA backend is used to compute scores for paired embeddings, which enables a similarity metric to be trained on potentially diverse situational datasets.
\subsection{Fine-tuning}
\vspace{1ex}
In general, x-Vector performance appears to be highly sensitive to both the amount and type of training data. This deep neural network has an extensive number of parameters, which for our solution totals 4.4 million parameters excluding the softmax output layer (it is not needed after training and will of course vary across different tasks). Any forensic dataset to be examined, in general, might not be very large in size due to the specialty of how the forensic dataset was collected in restricted acoustic environments. Additionally, speech samples may include variability due to vocal effort such as whisper-to-shout over 911 emergency calls, whereas others might include situational stress in a field location or interview room \cite{hansen2015speaker,campbell2009forensic}. Training the x-Vector model on a small or domain-mismatched dataset greatly affects the model's ability to generalize, often resulting in over-fitting, especially if the last few layers of the network are fully connected layers. Therefore, model adaptation with fine-tuning is indispensable in this case. More often in practice, existing networks trained on a large dataset such as VoxCeleb \cite{chung2018voxceleb2} would continue to be trained on a targeted smaller task-specific dataset. Given that if the small dataset is not drastically different in terms of context from the original training dataset, the pre-trained model is assumed to have already learned basic speaker-based features relevant to a target final task. There is a common practice to truncate the last layer (softmax layer) of the pre-trained network and replace it with a new softmax layer consistent with speaker labels of a new task to adapt the network. Since we expect pre-trained weights to be quite effective compared to randomly initialized weights, it is important not to modify them either too quickly or too much. Thus, an initial fine-tuning learning rate should be smaller than the one used for training from scratch. Additionally, it is beneficial to freeze the weights of the first few layers of the pre-trained network. Since the first few layers capture universal acoustic features that are also relevant to those for a new task, instead of keeping weights intact, we encourage the network to focus on learning task-specific features for the subsequent intermediate and high-level layers. 
\maketitle
\section{Domain Adversarial Training}
The commonly-used Fine-tuning method is generally effective in single-source domain adaptation, however, it cannot address the loss in performance resulting from a domain shift. Optimizing the classification objectives alone cannot guarantee effective generalization to multiple domains simultaneously without reducing the divergence between diverse distributions for each domain. Domain adversarial training utilizes domain information and promotes the emergence of features that are discriminative for speaker identity and invariant with respect to domain shift \cite{ganin2016domain,ganin2015unsupervised}.

We decompose a deep feed-forward architecture (see in Fig. 1) into three parts including ($i$) a universal feature extractor $G$ with the parameters $\theta_g$, ($ii$) a speaker label classifier $C_s$ with the parameters $\theta_s$, and a ($iii$) domain label classifier $C_d$ with parameters $\theta_d$. As training progresses, the parameters $\theta_g$ maximize the loss of the domain label classifier while simultaneously the parameters $\theta_d$ minimize the loss of the domain label classifier. Additionally, the parameters $\theta_s$ minimize the loss of the speaker label classifier. The multi-task softmax cross-entropy $\mathcal{J}$ loss can be written as,
\begin{eqnarray}
&Loss_{DAT}=\sum_{i=1}^{N} \sum_{j=1}^{M^i}\Big(\mathcal{J}\big(C_{s}\big(G(\vct x^{i}_j)\big), y^{i}_j\big)\nonumber\\
&-\lambda\mathcal{J}\big(C_{d}\big(G(\vct x^{i}_j)\big), d^{i}_j\big)\Big).
\end{eqnarray}

Given $N$ domains, here $(\vct{x}^i_j,y^i,d^i_j)$ represents the input data of the $j$-th utterance for domain $i$, the corresponding speaker label, and the domain label. Domain $i$ has $M^i$ utterances in total. The parameters $\theta_d$ minimize the domain classification loss while the parameters $\theta_s$ minimize the speaker classification loss. The parameters $\theta_g$ minimize the speaker classification loss and simultaneously maximize the domain classification loss, where the former makes the embeddings discriminative for speaker identity and the latter encourages domain-invariant embeddings to emerge in the course of optimization. The parameter $\lambda$ is a trade-off factor between the two losses. This process is implemented by a gradient reversal layer (GRL) \cite{ganin2015unsupervised}. The gradient reversal layer has no trainable parameters, and acts as an identity transform in the forward propagation. During the back-propagation, the GRL takes the gradient from the subsequent level, multiplies it by $-\lambda$ and passes it to the preceding layer.
\begin{figure}[h]
  \centering
  \includegraphics[width=1.0\linewidth]{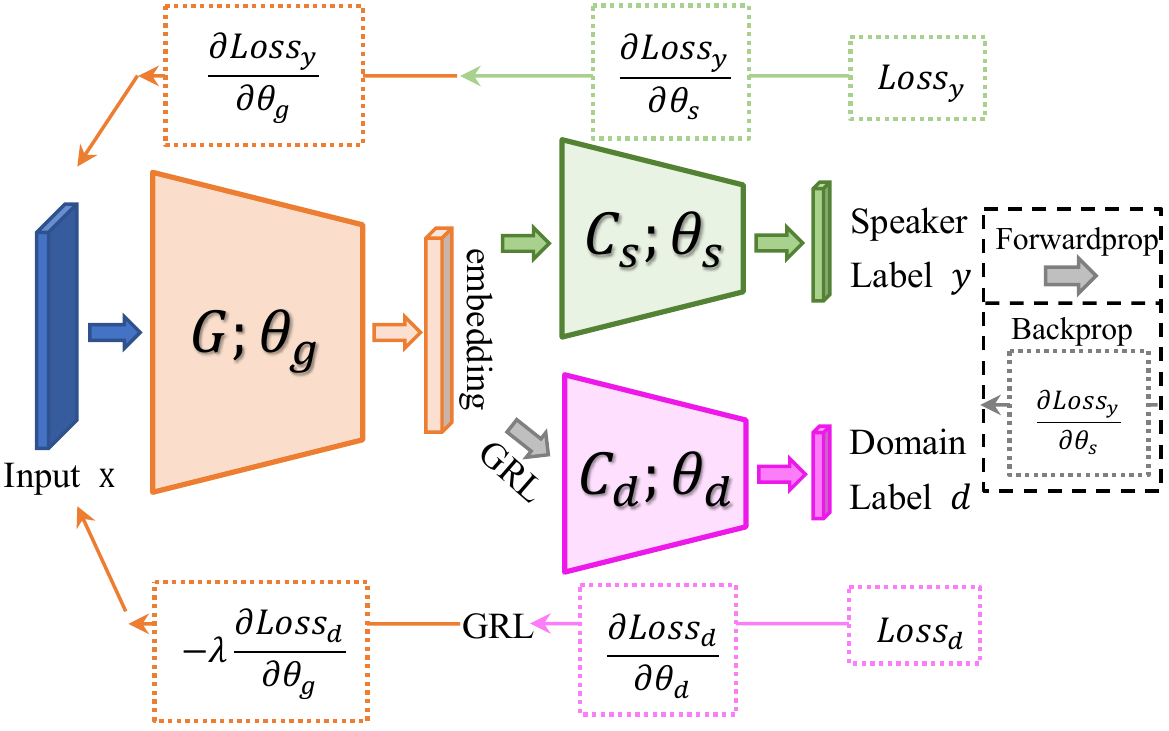}
  \captionsetup{font={footnotesize}}
  \caption{Domain adversarial training framework.}
  
\end{figure}
\maketitle
\section{Discrepancy Minimization on Features}
\vspace{1ex}
To mitigate the affect of domain discrepancy, multi-source domain adaptation based on discrepancy minimization, bridging the classification and discrepancy minimization, is employed to extract domain-invariant representations for all domains by means of respectively aligning the distributions of all domain pairs at both the frame-level and segment-level.
\subsection{Pair-wise Distribution Alignment}
Maximum mean discrepancy (MMD) is a pair-wise distribution discrepancy measure which is employed over a probability space by computing the mean squared difference of the statistics of samples \cite{gretton2006kernel,gretton2012kernel}.  Given that two generated distributions are identical, MMD assumes all corresponding statistics are the same and the distribution discrepancy will asymptotically equal 0. The definition in Eq. \eqref{eq:eq1} estimates the maximum mean discrepancy between each domain pair,
\setlength{\arraycolsep}{0.0em}
\begin{equation}
\mathcal{D}(\mathcal{X}_{a},\mathcal{X}_{b})=\left\|\mathbb{E}_{\vct x \sim \mathcal{X}_{a}}\mathit{\Phi}(\vct{x}^{a})-\mathbb{E}_{\vct x\sim \mathcal{X}_{b}}\mathit{\Phi}(\vct{x}^{b} )\right\|^{2},\label{eq:eq1}
\end{equation}
where MMD computes the mean square distance between the two collections $\vct{X}_{a} = \{\vct{x}_{i}^{a}\}_{i=1}^{\lvert\mathcal{X}_{a}\rvert}$ and $\vct{X}_{b} = \{\vct{x}_{j}^{b}\}_{j=1}^{\lvert\mathcal{X}_{b}\rvert}$ of $i.i.d.$ sampling from domain $\mathcal{X}_a$ and $\mathcal{X}_b$, $\mathit\Phi(\cdot)$ denotes a feature map, where the mapping is from acoustic frame-level features to an embedding space. Given $\lvert\vct{X}_{a}\rvert=L$ and $\lvert\vct{X}_{b}\rvert=M$, Eq. \eqref{eq:eq1} can be expanded as,
\setlength{\arraycolsep}{0.0em}
\begin{eqnarray}
&\mathcal{D}(\mathcal{X}_{a},\mathcal{X}_{b})=\frac{1}{L^2}\times\sum_{i=1}^{L}\sum_{i^\prime=1}^{L}\mathit{\Phi}(\vct{x}_{i}^{a})^\intercal\mathit{\Phi}(\vct{x}_{i^\prime}^{a}) \nonumber\\
&-\frac{2}{LM}\times\sum_{i=1}^{L}\sum_{j=1}^{M}\mathit{\Phi}(\vct{x}_{i}^{a})^\intercal\mathit{\Phi}(\vct{x}_{j}^{b}) \nonumber\\
&+\frac{1}{M^2}\times\sum_{j=1}^{M}\sum_{j^\prime=1}^{M}\mathit{\Phi}(\vct{x}_{j}^{b})^\intercal\mathit{\Phi}(\vct{x}_{j^\prime}^{b}).\label{eq:eq2}
\end{eqnarray}
\setlength{\arraycolsep}{5pt}

The dot product can be replaced with the kernel function $k(\cdot,\cdot)$,
\setlength{\arraycolsep}{0.0em}
\begin{eqnarray}
&\mathcal{D}(\mathcal{X}_{a},\mathcal{X}_{b})=\frac{1}{L^2}\times\sum_{i=1}^{L}\sum_{i^\prime=1}^{L}k(\vct{x}_{i}^{a},\vct{x}_{i^\prime}^{a})  \nonumber\\
&-\frac{2}{LM}\times\sum_{i=1}^{L}\sum_{j=1}^{M}k(\vct{x}_{i}^{a},\vct{x}_{j}^{b}) \nonumber\\
&+\frac{1}{M^2}\times\sum_{j=1}^{M}\sum_{j^\prime=1}^{M}k(\vct{x}_{j}^{b},\vct{x}_{j^\prime}^{b}).\label{eq:eq3}
\end{eqnarray}
\setlength{\arraycolsep}{5pt}
A widely-used kernel function is the radial basis function (RBF) kernel, which ensures that the MMD measure contains all moments of data in the feature space \cite{li2015generative}. This kernel function is written as,

\begin{equation}
k(\vct{x}^a,\vct{x}^b) = \rm{exp}(-\frac{1}{2{\sigma}^2}\left\|\vct{x}^a-\vct{x}^b\right\|^2),
\end{equation}
where $\sigma$ is a bandwidth parameter of the Gaussian kernel \cite{li2015generative}.

The motivation for a pair-wise distribution alignment is to ensure that the network predictions are consistent even if inputs are subject to an intrinsic/extrinsic \cite{hansen2015speaker} domain shift. As noted in \cite{long2015learning}, larger domain discrepancy gaps typically exist in deeper layers such as the fully-connected layer generating the embeddings. Therefore, we will minimize the discrepancy among embeddings produced by data from each domain. Additionally, the employed network uses an adaptive training scheme where samples are grouped into short segments (400 frames randomly out of 1000 frames) in each mini-batch, similar to the pre-training scheme based on sampling arbitrary-length segments from 200 to 400 frames. Moreover, the statistics pooling may also distort speech sequence features, especially based on temporal-related information. To alleviate this possible embedding distribution shift caused by speech sampling and statistics pooling, we also adapt frame-level features including the last TDNN layer's output before statistics pooling to avoid this inaccurate discrepancy estimate. Let $\mathbf{O}_a^l = \{{\mathbf{o}_i^l}\}_{i=1}^{\lvert\mathcal{X}^{a}\rvert}$ denote the collection of $l$-layer outputs from the distribution $\mathcal{O}_a^l$ for domain $\mathcal{X}_a$. Multiple domains will possess a domain shift with each other, where domain-invariant representations for each paired domain can be learned by minimizing the $Loss_{mmd}$ as follows,
\setlength{\arraycolsep}{0.0em}
\begin{eqnarray}
&Loss_{mmd}={\binom{N}{2}}^{-1}\sum_{i=1}^{N-1}\sum_{j=i+1}^N \Big(\mathcal{D}(\mathcal{O}_i^{\mathcal{F}_5},\mathcal{O}_j^{\mathcal{F}_5}) \nonumber\\
&+\mathcal{D}(\mathcal{O}_i^{{fc}_1},\mathcal{O}_j^{{fc}_1})\Big),
\end{eqnarray}
where $\mathcal{F}_5$ is the last TDNN layer before statistics pooling in the universal embedding extractor, and $fc_1$ is the first fully-connected layer which generates the embeddings. The discrepancy loss is computed between each pair of $N$ domains.
\subsection{Domain-specific Classification}
Each domain-specific subnet is followed by a softmax classifier. We use a softmax cross-entropy $\mathcal{J}$ loss for each classifier to ensure that the embedding distribution performance is improved for each domain. The classification loss function is written as,
\setlength{\arraycolsep}{0.0em}
\begin{equation}
Loss_{cls}=\sum_{i=1}^{N} \sum_{j=1}^{M^i}\mathcal{J}\big(C_{i}\big(G(\vct x^{i}_j)\big), y^{i}_j\big).
\end{equation}

Given $N$ domains, classification loss is computed for each domain-specific subnet. Here, $(\vct{x}^i,y^i)$ represents the acoustic frame-level input feature of an utterance for domain $i$ and the corresponding speaker label. Domain $i$ has $M^i$ utterances in total, and $G$ represents the universal embedding extractor, which maps the input feature to a universal embedding. Finally, $C_i$ is the classification subnet of the domain $i$ out of $N$ domains after employing the embedding extractor.
\subsection{Discrepancy Minimization Adaptation Framework}
Given an input data sequence, the universal feature extractor projects the sequence data into a temporal orderless embedding. The classification component will have four independent subnets corresponding to specific domains. The framework of the cross-domain adaptation in each step is illustrated in Fig. 2. 
Here, the multi-task loss function is formulated as,
\setlength{\arraycolsep}{0.0em}
\begin{equation}
Loss_{total} = \mu (Loss_{mmd}) + Loss_{cls}\label{eq:eq8},
\end{equation}
where $\mu$ is a variant adaptation factor with a progressive schedule from 0 to 1 in order to stabilize parameter sensitivity in the early adaptation stage.

Here, we employ our pre-trained x-Vector system as the universal embedding extractor, which is extended with classification subnets. The distributions of each domain are aligned simultaneously by minimizing domain discrepancies. Subsequently, the domain-invariant representations are specifically learned. Furthermore, the domain-specific classification subnets are employed to optimize recognition performance for individual domains of interest. 

\begin{figure}[h]
  \centering
  \includegraphics[width=1.0\linewidth]{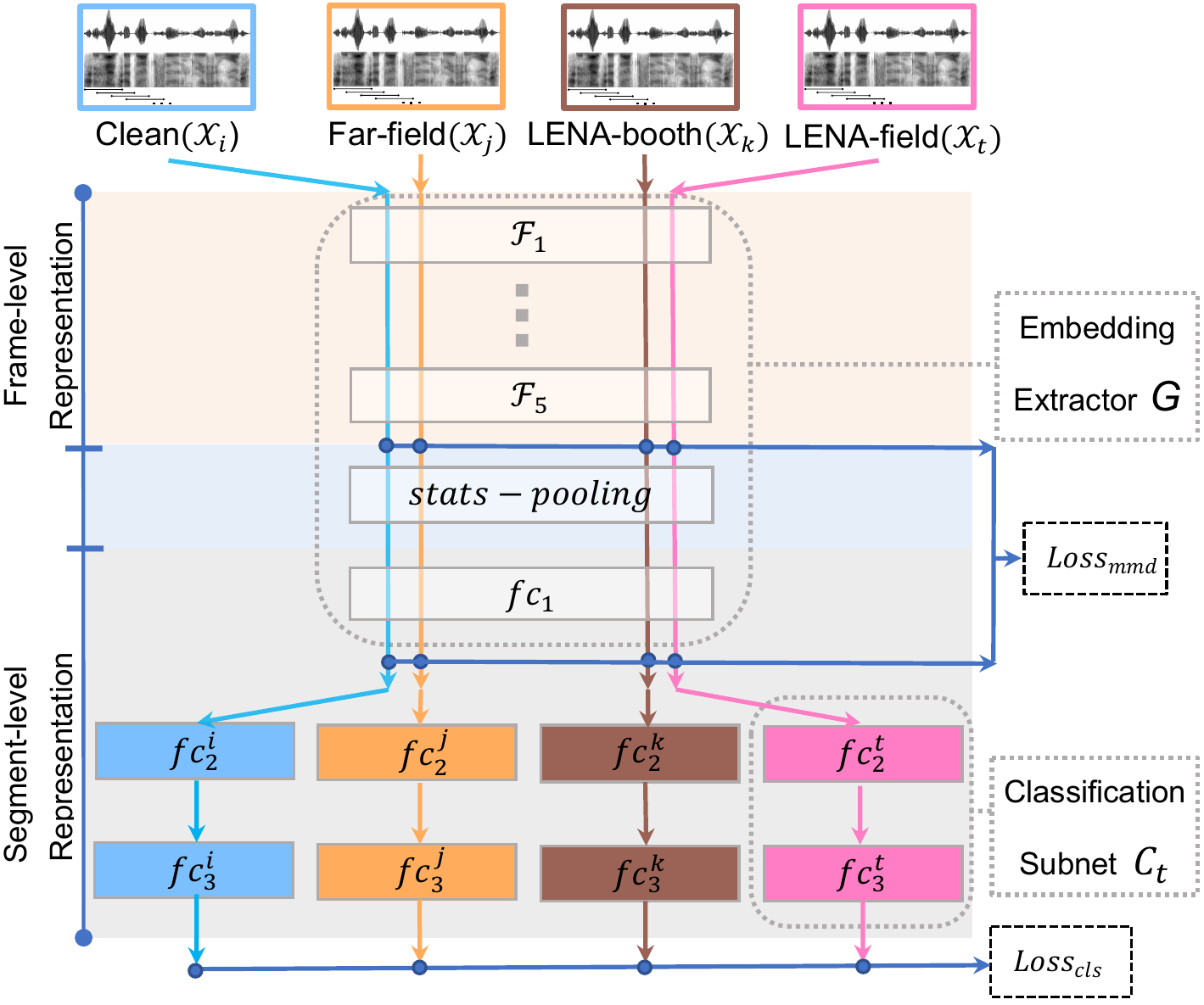}
  \captionsetup{font={footnotesize}}
  \caption{Discrepancy minimization adaptation framework.}
  
\end{figure}
\maketitle
\section{Moment Matching on The Classification Posterior}
\vspace{1ex}
For this study, four domains of our CRSS-Forensics dataset are investigated (details later in Sec. \uppercase\expandafter{\romannumeral5}). In general, multi-source domain data reduces the effectiveness of any single domain adaptation method. Additionally, the domain discrepancy will also vary in each pair of the domain-specific data. For this corpus, Clean, Far-field, and LENA-booth data were collected in the same recording environment (sound booth). For LENA-field data, speech data was collected using a portable LENA recording unit worn by the participant, with recording environments including seven pre-defined indoor and outdoor locations. Based on these corpus specifics, data collected in the sound booth have marginal domain shifts among each dataset (e.g. only close-talk mic (CTM) vs. desktop or distance mics at 4 ft and 8 ft). In contrast, a distinct domain discrepancy exists between LENA-field data and all data collected in the environment controlled sound booth. Therefore, we consider a multi-source domain adaptation based on moment matching and disposed multiple complex adversarial training procedures \cite{peng2019moment}. This method is employed to minimize the inter-domain discrepancies and transfer knowledge learned from sound-booth data to the more diverse LENA-field data by dynamically aligning moments of their feature distributions.
\subsection{Moment-matching Components}
Given labeled data collections $\vct{X}_1, \vct{X}_2, \dots, \vct{X}_N$ with distinct domain discrepancies for input feature space $\mathcal{X}$, the moment-matching process aims to find discriminative representations in the hypothesis embedding space $\mathcal{H}$, which minimizes the testing error on each domain. In \cite{peng2019moment}, the domain discrepancy was measured with the Moment Distance. Here, we use the measurement in Eq. \eqref{eq:eq1} instead to define the moment distance. The kernel function lifts the sample vectors into an infinite dimensional feature space and covers all orders of statistics, consequently minimizing MMD with this kernel which is equivalent to minimizing a distance between all moments of the two distributions \cite{gretton2006kernel}.

We employ a moment-matching model, which comprises of a universal embedding extractor $G$, along with a set of $N$ classifiers $\mathcal{C}=\lbrace\rm{C}_1, \rm{C}_2, \dots, \rm{C}_N\rbrace$ (e.g. this is the same setup as described in Sec. \uppercase\expandafter{\romannumeral3}). The MMD measurement minimizes the moment-related distance between domains as defined in Eq. \eqref{eq:eq1}. The overall loss function in Eq. \eqref{eq:eq8} is therefore rewritten as the following objective function,
\setlength{\arraycolsep}{0.0em}
\begin{eqnarray}
&\mathop{\min}_{G,\mathcal{C}}\sum_{i=1}^{N}\mathcal{J}_{\mathcal{X}_i}+\mu\mathop{\min}_{G}{\binom{N}{2}}^{-1}\sum_{i=1}^{N-1}\sum_{j=i+1}^N \nonumber\\
&(\mathcal{D}(\mathcal{O}_i^{\mathcal{F}_5},\mathcal{O}_j^{\mathcal{F}_5})+\mathcal{D}(\mathcal{O}_i^{{fc}_1},\mathcal{O}_j^{{fc}_1}))\label{eq:eq5},
\end{eqnarray}
where $\mathcal{J}_{\mathcal{X}_i}$ is a softmax cross-entropy loss for the classifier $\rm{C}_i$ for domain $\mathcal{X}_i$, and $\mu$ is a trade-off parameter with a progressive schedule. The objective of moment-matching adaptation is to match different distributions by minimizing the moment distance between multiple acoustic domains. Furthermore, for our task, we also intend to leverage knowledge learned from the noise-free sound-booth data to recalibrate the distribution for our diverse LENA-field data. 
\subsection{Adversarial Training Schema}
We follow the training paradigm suggested in \cite{saito2018maximum}, in order to utilize the domain-specific decision boundaries. Considering the relationship between class boundary and LENA-field samples, the paired domain-specific classifiers are taken as a discriminator to detect the presence of LENA-field samples from that reflecting the sound-booth domain. Paired classifiers are likely to classify those outliers in LENA-field samples differently. The tandem adaptive training includes the following three steps: 

\textbf{1)} Train the universal embedding extractor $G$ and classifier collection $\mathcal{C}$ to minimize moment distances as in Eq. \eqref{eq:eq1} among domains and perform classification on each domain. Model parameters are updated using the objective from Eq. \eqref{eq:eq5}. 

\textbf{2)} Fix the parameters of $G$, so as to maximize the discrepancies of classifier pairs. To measure the discrepancy of the two classifiers, we utilize the MMD as in Eq. \eqref{eq:eq1}, which formulates the objective in this training step,
\setlength{\arraycolsep}{0.0em}
\begin{eqnarray}
&\sum_{i=1}^{N-1}\big(\mathop{\min}_{{\rm{C}}_i}\mathcal{J}_{\mathcal{X}_i}-\mathcal{D}({{\rm{C}}_i}(\vct{X}_i),{{\rm{C}}_N}(\vct{X}_N))\big)+ \nonumber\\
&\mathop{\min}_{{\rm{C}}_N}\big(\mathcal{J}_{\mathcal{X}_N}-\frac{1}{N-1}\sum_{j=1}^{N-1}\mathcal{D}({{\rm{C}}_j}(\vct{X}_j),{{\rm{C}}_N}(\vct{X}_N))\big),\label{eq:eq11}
\end{eqnarray}
where ${\rm{C}}_i(\vct{X}_i)$ and ${\rm{C}}_N(\vct{X}_N)$ represent the probability outputs of ${\rm{C}}_i$ and ${\rm{C}}_N$ respectively from one of the sound-booth domains and LENA-field domains. The classification loss on each domain is added to stabilize system performance. 

\textbf{3)} Finally, we fix $\mathcal{C}$ and train $G$ to minimize the discrepancy of each classifier pair. The objective of this step is written as,
\setlength{\arraycolsep}{0.0em}
\begin{equation}
\mathop{\min}_{G}\sum_{i=1}^{N-1}\mathcal{D}({{\rm{C}}_i}(\vct{X}_i),{{\rm{C}}_N}(\vct{X}_N))\label{eq:eq12}.
\end{equation}
This entire procedure is summarized as Algorithm 1. For this solution, we train the classifiers and generator in an adversarial manner until the entire network (see in Fig. 3) reach a point of convergence.

\begin{figure}[h]
  \centering
  \includegraphics[width=1.0\linewidth]{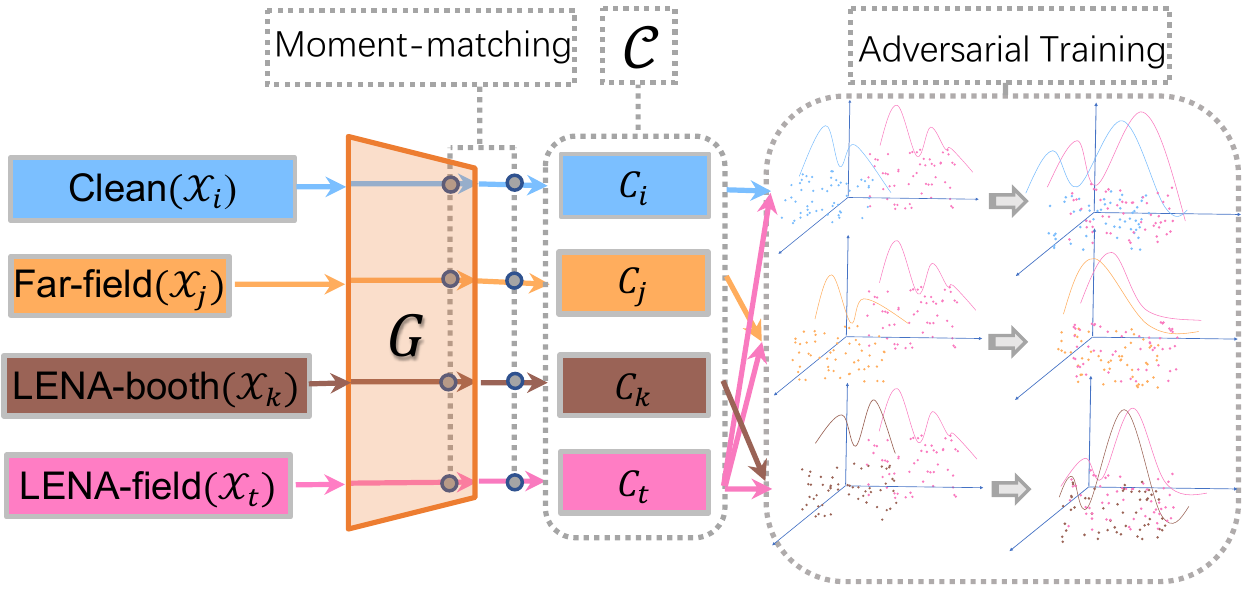}
  \captionsetup{font={footnotesize}}
  \caption{Moment-matching adaptation architecture.}

\end{figure}

\begin{algorithm}[h]
\captionsetup{font={footnotesize}}
  	\caption{ Moment-Matching Adaptation Network}
  	\label{alg:MMAN}
  	\begin{algorithmic}[1]
  	\begin{footnotesize}
  		\Require $\mathcal{X}=\lbrace\vct{X}_1,\vct{X}_2,\dots,\vct{X}_N\rbrace$, pre-trained $G$
  		\Ensure	pre-trained $G$ and a set of $N$ classifiers $\mathcal{C}=\lbrace\rm{C}_1, \rm{C}_2, \dots, \rm{C}_N\rbrace$
  		\State Given $T_1$ and $T_2$ training iterations 
  		\For{$t$ in $1:T_1$}
  		\For{$j$ in $1:N$}
  		\State Sample mini-batch $\lbrace\vct{x}_i^j,\vct{y}_i^j\rbrace_{i=1}^m$ from $\vct{X}_j$
  		\State Feed $\lbrace\vct{x}_i^j,\vct{y}_i^j\rbrace_{i=1}^m$ to $G$
  		\State Feed embeddings from $G$ to $\rm{C}_j$
  		\EndFor
  		\State \textbf{Update} $G$ and $\mathcal{C}$ according to Equation\eqref{eq:eq5}
  		\EndFor
  		\For{$t$ in $1:T_2$}
  		\For{$j$ in $1:N$}
  		\State Sample mini-batch $\lbrace\vct{x}_i^j,\vct{y}_i^j\rbrace_{i=1}^m$ from $\vct{X}_j$
  		\State Feed $\lbrace\vct{x}_i^j,\vct{y}_i^j\rbrace_{i=1}^m$ to $G$
  		\State Feed embeddings from $G$ to $\rm{C}_j$
  		\EndFor
        
        \State Fix $G$  
        
        \State \textbf{Update} $\mathcal{C}$ according to Equation\eqref{eq:eq11}
        
        \State Fix $\mathcal{C}$
        \For{$t^\prime$ in $1:4$}
        \State \textbf{Update} $G$ according to Equation\eqref{eq:eq12}
        \EndFor
  		\EndFor
  	\end{footnotesize}
  	\end{algorithmic}
\end{algorithm}
\maketitle
\section{Experiment}
\subsection{Data Description}
\subsubsection{\textbf{VoxCeleb}}
We use the Vox2 and Vox1 dev corpora for embedding training\cite{chung2018voxceleb2}, which is extracted from videos based on YouTube as training data for our pre-trained systems. Videos included in the dataset are recorded in a large number of challenging visual and auditory environments, including background conversations, laughter, overlapping speech, and varying room acoustics. Over 2.2 million utterances from $\approx$7300 speaker identities were used with corresponding annotation for speaker labels. Following a baseline Kaldi recipe, we use the dev and test splits from Vox2 and the dev split from Vox1 for embedding-oriented pre-training. 
\subsubsection{\textbf{CRSS-Forensic}}
As noted earlier, the CRSS-Forensics corpus\footnote{The CRSS-Forensics corpus will be released with a license.} contains read speech, prompted speech, and spontaneous speech in three conditions: clean (noise-free recorded in the sound booth), field recordings (with background noise and reverberation), and high stress (actual police interviews).  Two phases are included in the recording process. Phase-1 contains speech data recorded in a controlled noise-free sound booth, and diverse public acoustic field environments. Speech in Phase-2 is collected in a law enforcement facility, using an interview room with an actual police officer/detective. Figure 5 shows sample recording environments for the noise-free sound booth, public field environments, and police interview room.

For the sound booth in Phase-1, speech data is simultaneously recorded using multiple wired/wireless microphones (sample rate: 44.1 kHz) and a participant body-worn mobile data collection platform called LENA unit (sample rate: 16 kHz). Microphones are positioned at 4 different locations in a $13^\prime\times13^\prime$ sound booth (SB); one being a close talk microphone (CTM), the other three representing distances from each microphone to the speaker are 18-24 in (Desktop Mic (DTM)), 4 ft, and 8 ft, as shown in Fig. 4. In LENA-field environments, speech is collected by a LENA unit worn by the participant, with recording environments that include seven indoor and outdoor locations (7IOL): office, hallway, cafeteria, parking lot, game room, lobby, walking path (see in Fig.5 (a-g)).
In Phase-2, speech data is simultaneously recorded using a participant body-worn LENA unit and a microphone. Here a detective conducts an investigative interview of the participant concerning a specific scenario while following standard procedures in law enforcement interview room (LEIR) (see in Fig. 5 (h)). 

\begin{table}[h]
\renewcommand{\arraystretch}{1.3}
\captionsetup{font={footnotesize}}
 \caption{Data Statistics for CRSS-Forensics Corpus}
 \label{data_stat}
  \centering

  \vspace{-1.5ex}
    \begin{tabular}{cccc}
    \hline
    \hline 
    \multicolumn{1}{c}{Info} & Session Name & Duration & Speaker \\
    \hline 
    
    \multirow{3}{*}{Phase-1.SB} & Clean (CTM \& DTM) & 32 h/channel  & 75    \\ 
         								   & Far-field (4 ft \& 8 ft) & 32 h/channel  & 75     \\
                                           & LENA-booth & 33.9 h & 75     \\
                       \hline
                       Phase-1.7IOL & LENA-field & 99.4 h & 75     \\
    \hline
    Phase-2.LEIR   & Interview (LENA \& Mic) & 20.4 h & 58     \\
    \hline
    \hline 
    \end{tabular}%
  
  \vspace{-1.5ex}
\end{table}%

Table $\rm\uppercase\expandafter{\romannumeral1}$ summarizes the specific acoustic data size for each session. For the 75 speakers in the corpus, 65 are native English speakers and 10 non-native speakers, with 27 male speakers and 48 female speakers. Each participant was allowed to opt-out of Phase-2 (i.e., IRB protocol due to high-stress level exposure), so there are 17 speakers absent from the Phase-2 police interview set. 

In this study, data from Phase-1 is used for multi-source domain adaptation. We note that various recording environments are considered as the extrinsic characteristics for audio samples, while speech from speakers under stress for the Interview in Phase-2 contain intrinsic variations. Consequently, data in Phase-2 is not compatible with environmental mismatch data in Phase-1 for domain-invariant information extraction. We consider 16 speakers (8 male, 8 female) out of 75 speakers for each set to perform evaluation. There exists no speaker overlap between the training and evaluation sets, abiding by an open-set protocol. The number of test trials is over 22,000 total. For sound booth data, speech data with CTM and DTM are designated as the Clean set; data collected from the remaining two distant mics (4 ft \& 8 ft) are taken as Far-field data; and data recorded by the LENA body-worn unit was used to explore channel mismatch influence.

\begin{figure*}[ht]
  \centering
  \vspace{-1.5ex}
  \includegraphics[scale=0.7]{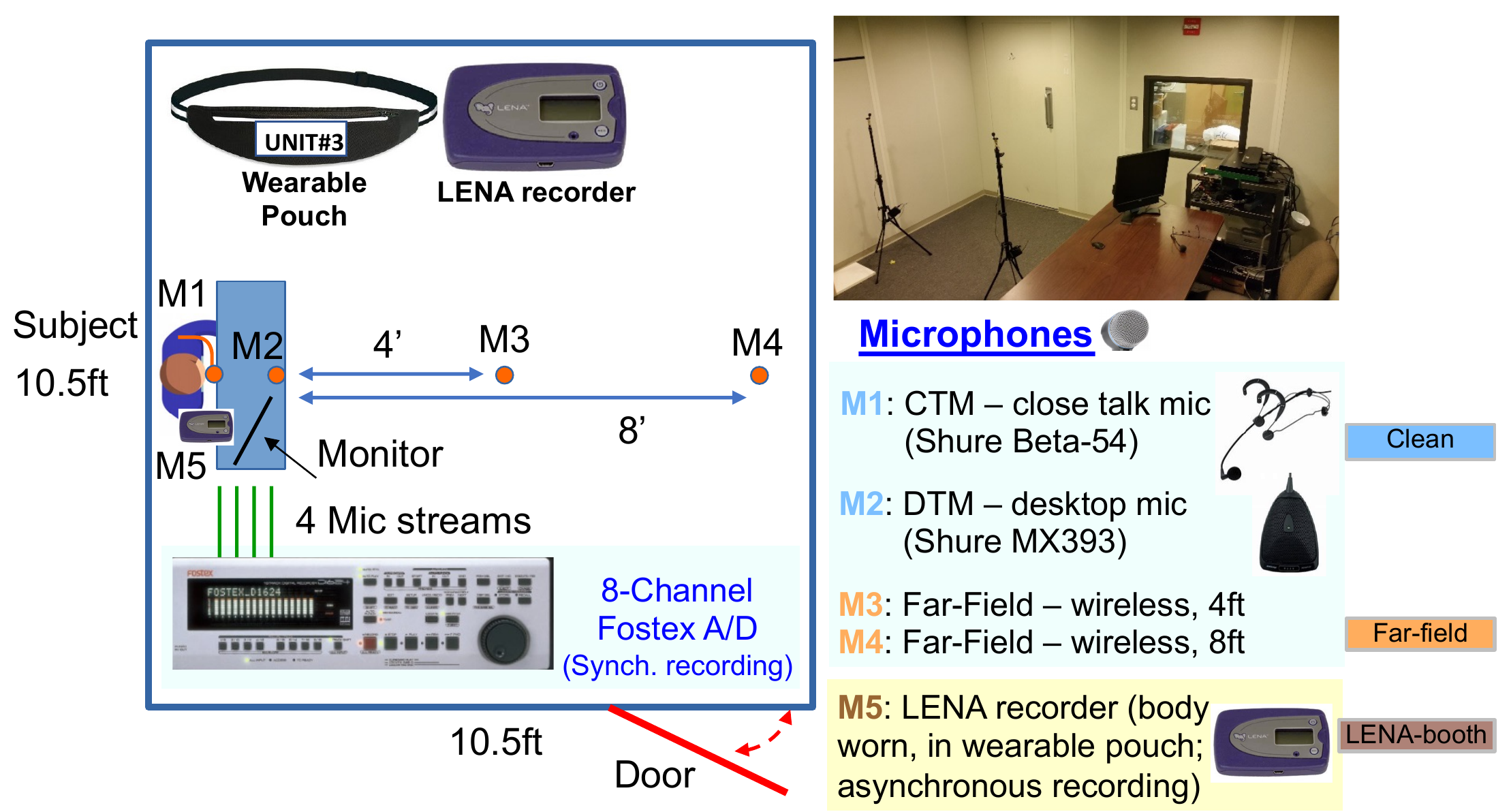}
  \captionsetup{font={footnotesize}}
  \caption{Sound booth forensic voice data collection setup, including 5 audio streams (M1: close-talk mic, M2: desk-top mic, M3 \& M4: far-field distance mics, M5: asynchronous body-worn LENA recorder).}

\end{figure*}

\begin{figure*}[ht]
  \centering
  \vspace{-1ex}
  \includegraphics[width=0.99\linewidth]{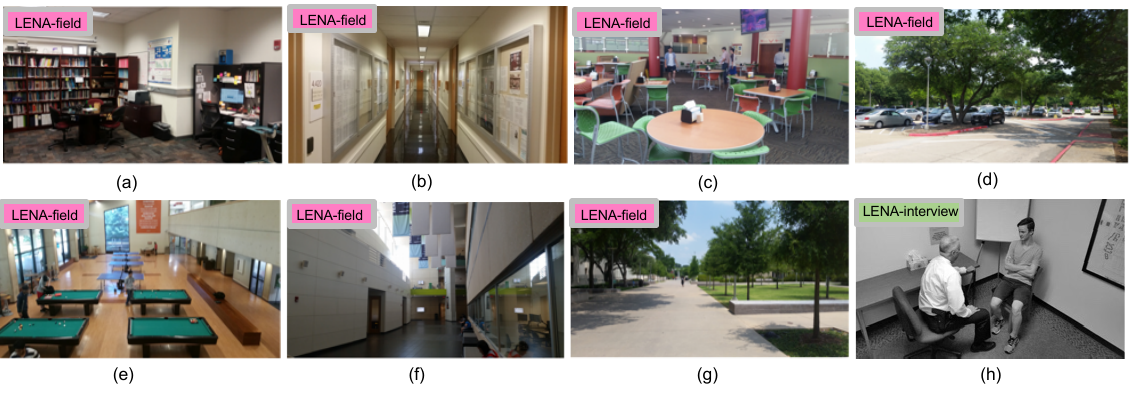}
  \captionsetup{font={footnotesize}}
  \caption{LENA-Field forensic audio collection locations (a) to (g); plus Police Interview location (LENA-Police) recording environments (h).}
  
  \vspace{-1.5ex}
\end{figure*}
\subsection{Pre-training System Setup}

Gender independent i-Vector extractors were trained on the VoxCeleb dataset to produce 400-dimensional i-Vectors. 20-dimensional MFCCs were then augmented with their delta and double-delta coefficients, producing a set of 60-dimensional MFCC feature vectors. 

In order to implement a competitive and fair baseline, we developed the x-Vector system. Our model is similar to an x-Vector Kaldi recipe\footnote{\url{https://kaldi-asr.org/models/m7}} with respect to VoxCeleb corpora and network architecture. The model architecture consists of 5 time-delay layers, which model temporal context information, followed by a statistical pooling layer to map into a fixed-dimensional vector at the segment-level. This is followed by two fully-connected layers with 512 units in each layer and the probability output layer. We extract 30-dimensional MFCC features using a frame width of 25ms and window shift of 10ms. Training data is augmented with noise, music, and babble speech from the MUSAN corpus \cite{snyder2015musan}, and reverberation of the RIR NOISES\footnote{\url{http://www.openslr.org/28}} corpus. The augmented data consist of 7323 speakers and 2.2M utterances. Specially, all utterances shorter than 4 seconds in duration, and all speakers with fewer than 8 utterances are set aside in the data pre-processing phase. Cepstral mean normalization with a sliding window of 3 seconds was employed to suppress channel effects. We use an Adam optimizer with betas of (0.9, 0.98) to update model parameters, initializing the learning rate of 1e-3. The learning rate was adjusted with the warm-up scheduling named “Noam” in \cite{vaswani2017attention}. Batch normalization and Dropout are also used to perform regularization at each layer. Finally, a mini-batch of 32 samples is used at each iteration.
\subsection{Fine-tuning Setup}
In this work, all adaptation methods only update parameters of the last time-delay layer before statistics pooling and the first fully-connected layer after pooling in the pre-trained x-Vector model. The last fully-connected layer is replaced according to speaker labels from the CRSS-Forensic data. We perform fine-tuning on the pre-trained model with our data using the Adam optimizer to retrain the model for 40 epochs using a batchsize of 64. The learning rate is scheduled using the formula,
\begin{equation}
\eta_p = \frac{\eta_0}{(1+\alpha p)^\beta},
\end{equation}
where $\eta_0$ = 1e-4, $\alpha$ = 10, $\beta$ = 0.75 and $p$ is set to linearly increase from 0 to 1 corresponding to the training steps.
\subsection{Domain Adversarial Training System Setup}

We keep the pre-trained x-Vector model as the feature extractor, and extract embeddings from the first fully-connected layer after statistic pooling. For the speaker label classifier, we retain the three fully-connected layers ($Embedding$ $\rightarrow$ 512 $\rightarrow$ 512 $\rightarrow$ 59), and use a simpler architecture ($Embedding$ $\rightarrow$ GRL (described in Sec. \uppercase\expandafter{\romannumeral3}) $\rightarrow$ 128 $\rightarrow$ 4) for domain classification. The model is trained on 64-sized batches. In order to suppress noisy signals from the domain classifier at the early training stages instead of fixing the trade-off factor $\lambda$, we gradually change this value from 0 to 1 using the following schedule:
\begin{equation}
\mu = \frac{2}{1+\rm{exp}(-\theta p)} - 1\label{eq:eq_progress},
\end{equation}
where $\theta$ = 10, $p$ is set to linearly increase from 0 to 1 corresponding to the training steps.
\subsection{Discrepancy-minimizing System Setup}

In addition to the pre-trained model, we keep the first fully-connected layer after statistics pooling as part of the embedding extractor. Embeddings from each domain are processed by a fully-connected layer with 512 units and the final layer which outputs speaker posterior probabilities, respectively. The discrepancy-minimizing model is trained on the CRSS-Forensic data for 40 epochs using a batchsize of 64. We use the Adam optimizer to update parameters of the partial pre-trained model with a learning rate of 1e-4, and for each classification subnet, the learning rate is set to 1e-3. Since there exists no parameter-wise differences between each subnet in the early adaptation stage, Eq.\eqref{eq:eq8} may result in noisy activations. To stabilize parameter sensitivity, a progress strategy \cite{ganin2015unsupervised} is used for Eq.\eqref{eq:eq8} as noted in Eq.\eqref{eq:eq_progress}.

\subsection{Moment-matching System Setup}
The model setup for moment-matching system is basically consistent with the previous description for the discrepancy-minimizing system with the exception for the training schema and loss functions constituted by another discrepancy measurement (details in Sec. \uppercase\expandafter{\romannumeral5}). The moment-matching model is trained for 40 epochs as $T_1$ in Algorithm \ref{alg:MMAN} then proceeds for another 40 epochs as $T_2$ in Algorithm \ref{alg:MMAN} with a batchsize of 64. 

For each system, we take embeddings from the outputs of the first fully-connected layer after statistics pooling for evaluation purposes, and score trials using PLDA \cite{prince2007probabilistic} after performing dimensionality reduction to 200 using LDA and length-normalization. Here, LDA and PLDA multi-conditional training are conducted in each system with generated embeddings from the training portion in CRSS-Forensic corpus containing 59 speakers, which can compensate for domain mismatch.
\section{Result and Analysis}
\begin{figure*}[ht]
\centering
\vspace{-1.5ex}
\subfigure[Clean ${\textcolor[HTML]{3399FF}{\blacksquare}}$]{\includegraphics[width=0.449\linewidth]{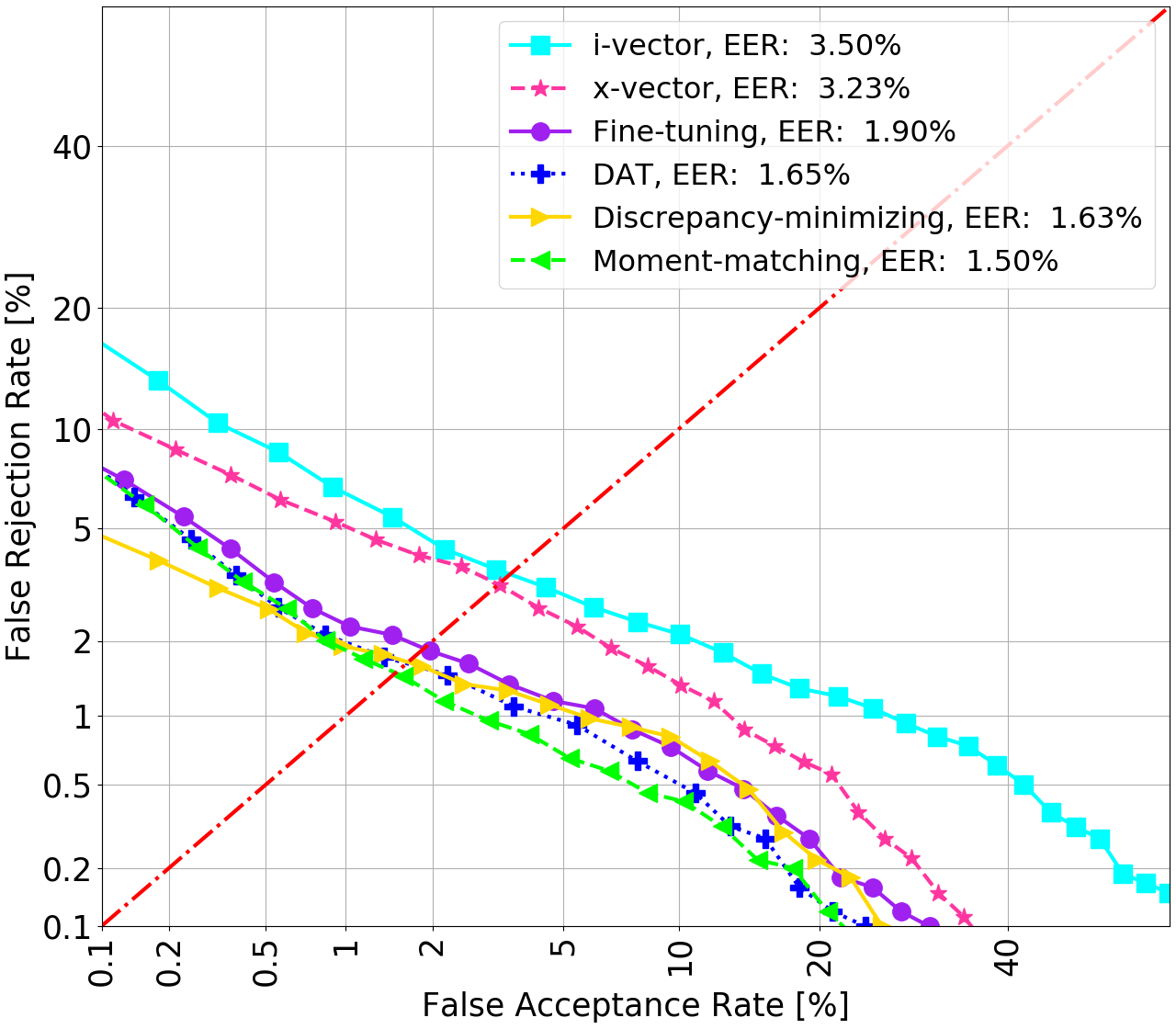}}
\subfigure[Far-field ${\textcolor[RGB]{255,125,10}{\blacksquare}}$]{\includegraphics[width=0.449\linewidth]{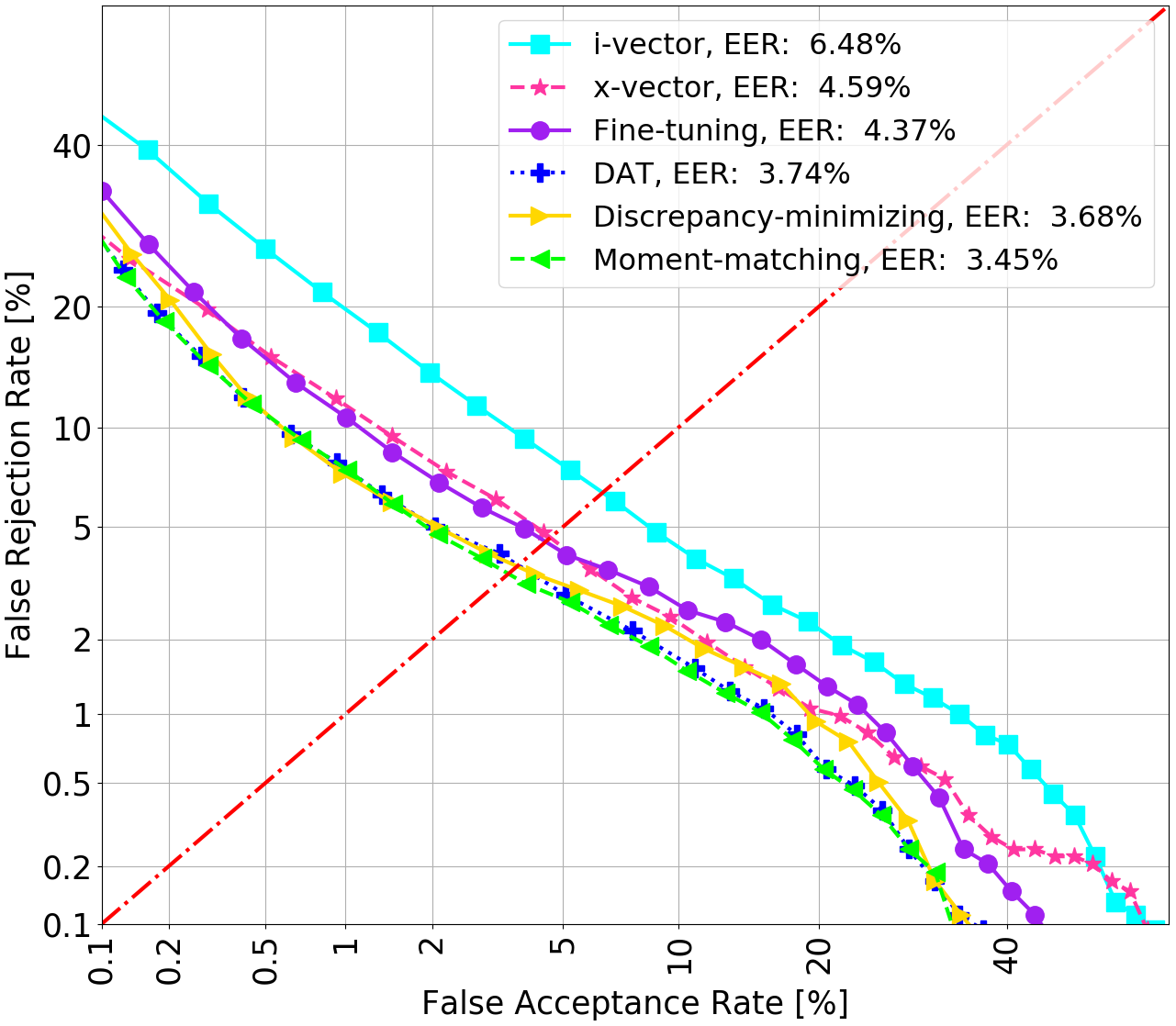}}

\subfigure[LENA-booth ${\textcolor[RGB]{96,17,0}{\blacksquare}}$]{\includegraphics[width=0.449\linewidth]{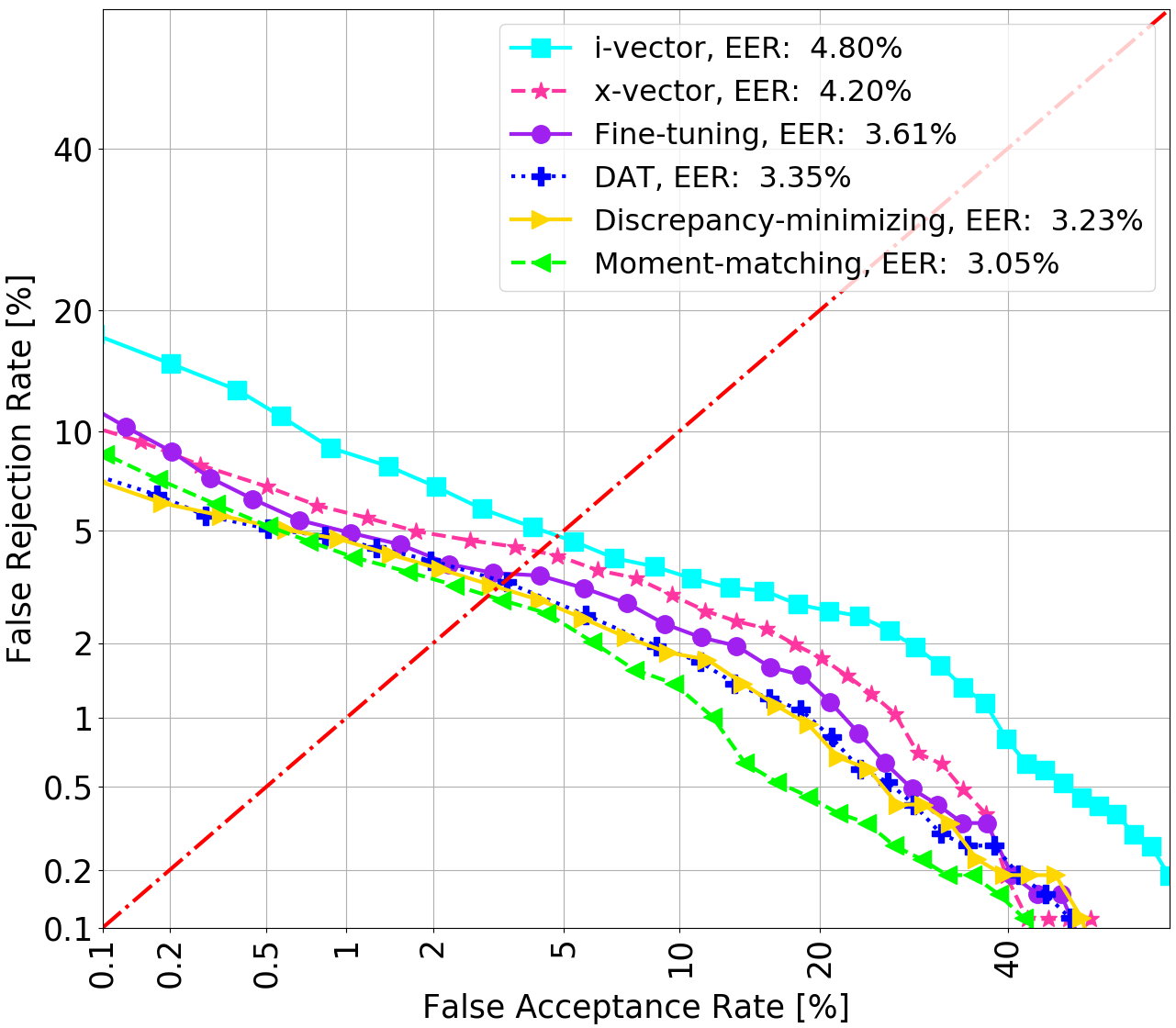}}
\subfigure[LENA-field ${\textcolor[RGB]{255,54,160}{\blacksquare}}$]{\includegraphics[width=0.449\linewidth]{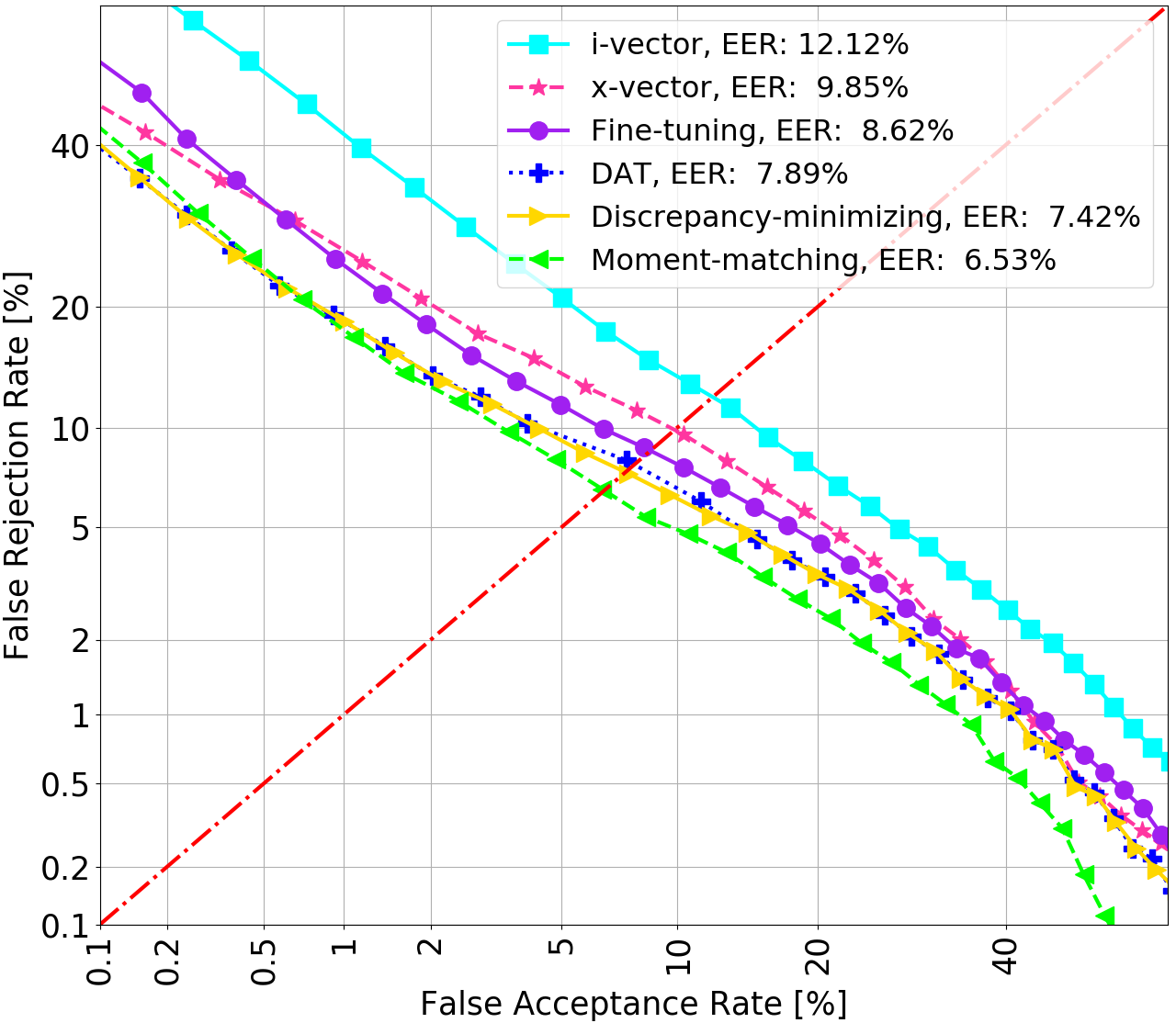}}
\captionsetup{font={footnotesize}}
\caption{DET curves for each system on four domains.}
\label{fig:pictures2} 
\vspace{-2ex}
\end{figure*}
This section focuses on the analysis of each system implemented based on setups described in Sec. \uppercase\expandafter{\romannumeral6}. To evaluate these experiments, we adopt several measurement criteria concentrating on evaluating discrimination abilities and calibration of speaker recognition systems. In terms of speaker recognition system evaluation, the trade-off between missed speakers $P_{miss}$ and false alarms $P_{FA}$ has always been a key diagnostic tool. The Detection Error Trade-off (DET) curve \cite{Martin1997TheDC} reflects what happens as the decision threshold is swept across the entire operating range. Noting that $P_{miss}$ and $P_{FA}$ move in opposite directions as the decision threshold is shifted, a point where $P_{miss} = P_{FA}$ called the Equal Error Rate (EER), provides a standard point for the discrimination capability of the system. However, the EER does not measure calibration (the ability to set decision thresholds). It is noted that the recognition system actually produces the log-likelihood-ratio $\mathcal{L}_t$ of the score for each trial. The Log-likelihood-ratio cost function \cite{brummer2006application} $C_{llr}$ is a simultaneous measurement of the discrimination abilities of the log-likelihood-ratio scores and calibration for application-independent detection decisions, which is formulated as,
\setlength{\arraycolsep}{0.0em}
\begin{eqnarray}
&C_{llr}(\mathcal{L}_t)=\frac{1}{2}\Big(\frac{1}{N_{tar}}\sum_{t\in \rm{T}_{tar}}\rm{log}(1+e^{-\mathcal{L}_t})\nonumber\\ 
&+\frac{1}{{N}_{non}}\sum_{t \in \rm{T}_{non}}\rm{log}(1+e^{\mathcal{L}_t})\Big),
\end{eqnarray}
where $\mathcal{L}_t$ is the log-likelihood-ratio score of trial $t$, $\rm{T}_{tar}$ is a set of $N_{tar}$ target trials and $\rm{T}_{non}$ is a set of $N_{non}$ non-target trials. Furthermore, $C_{llr}$ can be minimized as measured on a warped log-likelihood-ratio score $\mathcal{L}^\prime_t=w(\mathcal{L}_t)$ scale using a monotonic rising warping function $w$, resulting in the performance measure $C_{llr}^{min}=C_{llr}\big(\{w(\mathcal{L}_t)\}\big)$. Finally, $w$ can be optimized using the Pool Adjacent Violators (PAV) approach \cite{brummer2006application}. We visualized each speaker recognition system performance on each acoustic domain using DET curves (see in Fig. 6).

\subsection{Score Calibration}
\begin{table*}[ht]
  \centering
  \renewcommand{\arraystretch}{1.3}
  \captionsetup{font={footnotesize}}
  \caption{Score Calibration Result for Each System on Four Domains}
  \vspace{-1ex}
    \begin{tabular}{cccc|ccc|ccc|ccc}
    \hline
    \hline 
     &\multicolumn{3}{c}{Clean}&\multicolumn{3}{c}{Far-field}&\multicolumn{3}{c}{LENA-booth}&\multicolumn{3}{c}{LENA-field}\\
     
     \hline
     \specialrule{0em}{1pt}{1pt}
     & $C_{llr}$ &$C_{llr}^\prime$ & $C_{llr}^{min}$ & $C_{llr}$ &$C_{llr}^\prime$ & $C_{llr}^{min}$& $C_{llr}$ &$C_{llr}^\prime$ & $C_{llr}^{min}$& $C_{llr}$ & $C_{llr}^\prime$ & $C_{llr}^{min}$   \\
     \specialrule{0em}{1pt}{1pt}
      \hline 
    i-Vector  & 0.241 & 0.172& 0.138& 0.335 & 0.247& 0.231& 0.364 & 0.234& 0.176& 0.591 & 0.428& 0.399\\ 
    
    x-Vector  & 0.211& 0.130& 0.110& 0.323 & 0.189& 0.168& 0.331 & 0.180& 0.136& 0.610 & 0.315& 0.314\\
    Fine-tuning  & 0.200 & 0.098& 0.072& 0.364 & 0.183& 0.167& 0.288 & 0.144& 0.123& 0.621 & 0.323& 0.293\\
    DAT  & 0.272 & 0.118& 0.064& 0.312 & 0.145& 0.134& 0.294 & 0.114& 0.105& 0.528 & 0.276& 0.254\\
    Discrepancy-minimizing  & 0.172 & 0.085& 0.063& 0.298 & 0.155& 0.139& 0.231 & 0.124& 0.109& 0.526 & 0.284& 0.256\\
    Moment-matching  & 0.272 & 0.080& 0.060& 0.276 & 0.140& 0.131& 0.211 & 0.111& 0.100& 0.479 & 0.254& 0.233\\

    \hline
    \hline 
    \end{tabular}%
  
  \vspace{-1ex}
\end{table*}%
The recognition system produces the log likelihood ratio in terms of PLDA scores, but the scores are uncalibrated which may adversely affect the validity and reliability of this evaluation. Score calibration has been recognized as an important component in effective evaluation of current speaker recognition systems \cite{kelly2016score,brummer2006application,van2007introduction}. Thus, we calibrate the log-likelihood-ratio (LLR) scores by finding a linear transform that optimizes the CLLR measure to reach a value of $C_{llr}^\prime$ closer to $C_{llr}^{min}$. In this study, we employ a commonly-used linear calibration transformation,
\begin{equation}
s^\prime=w_0+w_1s,
\end{equation}
where an uncalibrated score $s$ is transformed into a calibrated score $s^\prime$ using offset $w_0$ and scaling factor $w_1$ parameters. Logistic regression optimization \cite{pigeon2000applying} is employed to acquire the two calibration parameters $w_0$ and $w_1$. We summarized score calibration results for each speaker recognition system on each acoustic domain using DET curves (as shown in Table $\rm\uppercase\expandafter{\romannumeral2}$), where $C_{llr}^\prime$ corresponds to $C_{llr}$ after calibration.

\subsection{Location Analysis}

Next, speaker recognition (SR) performance is assessed in terms of EER and $C_{llr}$ for both i-Vector and x-Vector system over the range of evaluation datasets (as shown in Table $\rm\uppercase\expandafter{\romannumeral3}$). The pre-trained x-Vector system shows better speaker recognition performance, so it is taken as the embedding extractor for subsequent adaptation methods. Additionally, it is noted that the x-Vector architecture is based on a deep neural network, which allows for fine-tuning and to concatenate with other deep-learning structures. In terms of impact due to domain mismatch on system performance, channel, speaker-to-mic distance, and environmental noise all exert some mismatch influence on system recognition performance, with noise mismatch having the greatest impact. 

\begin{table}[ht]
  \centering
  \renewcommand{\arraystretch}{1.3}
  \captionsetup{font={footnotesize}}
  \caption{SR Result for i-Vector \& x-Vector Systems}
  \vspace{-1ex}
    \begin{tabular}{cccc|ccc}
    \hline
    \hline 
     &\multicolumn{3}{c}{i-Vector}&\multicolumn{3}{c}{x-Vector}\\
     
     \hline
     \specialrule{0em}{1pt}{1pt}
     & EER & $C_{llr}^\prime$ & $C_{llr}^{min}$ & EER & $C_{llr}^\prime$ & $C_{llr}^{min}$  \\
     \specialrule{0em}{1pt}{1pt}
      \hline 
    Clean & 3.50\%  & 0.172   & 0.138 & 3.23\% & 0.130& 0.110 \\ 
    
    Far-field & 6.48\%  & 0.247   & 0.231 & 4.59\% & 0.189& 0.168 \\
    LENA-booth & 4.80\%  & 0.234   & 0.176 & 4.20\% & 0.180& 0.136 \\
    LENA-field & 12.12\%  & 0.428   & 0.399 & 9.85\% & 0.315& 0.314 \\

    \hline
    \hline 
    \end{tabular}%
  
  \vspace{-1ex}
\end{table}%

\begin{table}[h]
  \centering
  \renewcommand{\arraystretch}{1.3}
  \captionsetup{font={footnotesize}}
  \caption{SR Result for x-Vector System in 7 Locations}
  \vspace{-1ex}
    \begin{tabular}{ccccc}
    \hline
    \hline 
     7IOL &EER&$C_{llr}$&$C_{llr}^\prime$&$C_{llr}^{min}$\\
     \specialrule{0em}{1pt}{1pt}
     \hline
     Cafeteria&12.61\%&0.807&0.409&0.372\\
     Game room&13.78\%&0.900&0.444&0.416\\
     Hallway&9.23\%&0.632&0.314&0.292\\
     Lobby&8.55\%&0.493&0.293&0.267\\
     Office&6.64\%&0.395&0.241&0.224\\
     Parking Lot&7.35\%&0.401&0.262&0.240\\
     Walking Path&9.01\%&0.545&0.311&0.290\\
     
    \hline
    \hline 
    \end{tabular}%
  
  \vspace{-1ex}
\end{table}%

The LENA-field set includes 7 naturalistic locations (7IOL as shown in Fig. 5). We evaluate x-Vector system performance on data across each environmental location (as shown in Table $\rm\uppercase\expandafter{\romannumeral4}$). Results show that speech data captured in public cafeteria and game room locations had the lowest speaker recognition results versus other locations. Cafeteria and game room data contain secondary people talking, and sporadic background music and random noise/sound events which are clearly heard especially in the game room. Speaker identity is more easily discriminated with data from the office context, since noise content is less, though background talking can occur at times. Other locations contain varying amounts of reverberation and ambient noise, also resulting in degradation in recognition performance.

\subsection{Fine-tunning Layers Selection}

In order to explore the best fine-tuning result, we performed fine-tuning of the pre-trained x-Vector system for different layers. Here, we present the fine-tuning results across 3 different options: ($\mathcal{F}_4,\mathcal{F}_5,fc_1$), ($\mathcal{F}_5,fc_1$) and $fc_1$ (definition see Fig. 2) as shown in Table $\rm\uppercase\expandafter{\romannumeral5}$.

\begin{table}[ht]
  \centering
  \renewcommand{\arraystretch}{1.3}
  \captionsetup{font={footnotesize}}
  \caption{$C_{llr}$ Result for x-Vector Model Fine-tuning}
  \vspace{-1ex}
    \begin{tabular}{ccc|cc|cc}
    \hline
    \hline 
     &\multicolumn{2}{c}{$\mathcal{F}_4,\mathcal{F}_5,fc_1$}&\multicolumn{2}{c}{$\mathcal{F}_5,fc_1$}&\multicolumn{2}{c}{$fc_1$}\\
     \hline 
     \specialrule{0em}{1pt}{1pt}
     &$C_{llr}^\prime$ & $C_{llr}^{min}$ & $C_{llr}^\prime$ & $C_{llr}^{min}$ & $C_{llr}^\prime$ & $C_{llr}^{min}$\\
     \specialrule{0em}{1pt}{1pt}
      \hline 
    Clean & 0.102  & 0.075   & 0.098 & 0.072 & 0.100& 0.074 \\ 
    
    Far-field & 0.174  & 0.191   & 0.183 & 167 & 0.187& 0.170 \\
    LENA-booth & 0.150  & 0.127   & 0.144 & 0.123 & 0.147& 0.125 \\
    LENA-field & 0.334  & 0.304   & 0.323 & 0.293 & 0.328& 0.298 \\

    \hline
    \hline 
    \end{tabular}%
 
  \vspace{-1ex}
\end{table}%
Table $\rm\uppercase\expandafter{\romannumeral5}$ shows that fine-tuning of the pre-trained x-Vector model achieves the best result for $C_{llr}^\prime$ and $C_{llr}^{min}$ when applied in the last layer before statistics pooling, and the first layer after pooling. By performing fine-tuning in the proper layers, knowledge of the pre-training data is effectively transferred towards the current model, and the model also learns effective speaker information for the new dataset. We only fine-tune subsequent layers to maintain the learned universal speaker features from undue distortion. The fine-tuned system lowers pre-trained system's EER with a relative decrease of 41.18\%, 4.79\%, 12.49\%, 14.05\% in each set, respectively, with an averaged relative decrease in EER of 18.13\%. Obviously, the Clean set benefits the most from fine-tuning.
\subsection{Adaptive Training with Domain Information}
\begin{figure*}[ht]
  \centering
  \vspace{-2ex}
  \includegraphics[width=0.99\linewidth]{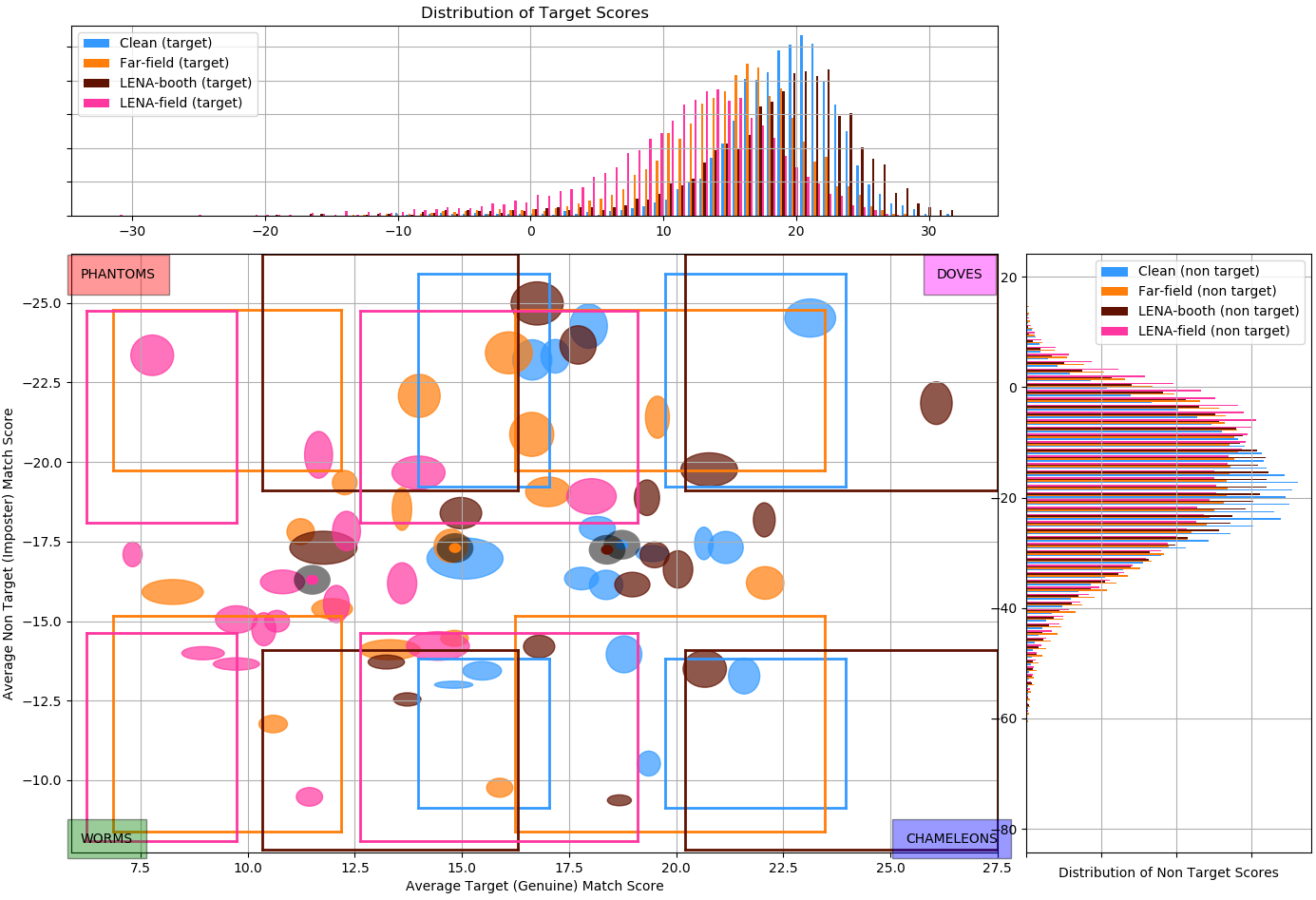}
  \captionsetup{font={footnotesize}}
  \caption{Zoo-plot for discrepancy minimization system.}
  
  \vspace{-1.5ex}
\end{figure*}
Fine-tuning effectively improves speaker recognition performance for the pre-trained x-Vector system, though the achieved improvement is unbalanced across each domain. Therefore, it is necessary to explore other options to improve use of domain information for better speaker recognition performance. This can be achieved by DAT \cite{ganin2016domain} which uses a gradient reversal layer to remove the domain variation and projects the alternate domain data towards the same subspace. This approach learns a domain-invariant and speaker-discriminative feature representation. The DAT system lowers EER of the pre-trained system with a relative decrease of 48.92\%, 18.52\%, 20.24\%, 19.90\% for each set, respectively, with an averaged relative EER decrease of 26.90\%. 

\subsection{Frame-level Effectiveness in Discrepancy Minimization}

An alternative strategy to address mismatch is to consider a multi-task loss function in order to improve the pre-trained system performance across acoustic environments. The main objective is to minimize the domain discrepancy so as to achieve system performance gains for each domain. 
\begin{table}[h]
  \centering
  \renewcommand{\arraystretch}{1.3}
  \captionsetup{font={footnotesize}}
  \caption{SR Result for Discrepancy Minimization System}
  \vspace{-1ex}
    \begin{tabular}{cccc|cccc}
    \hline
    \hline 
     &\multicolumn{3}{c}{frame-level}&\multicolumn{3}{c}{frame-level \& seg-level}\\
     \hline
     \specialrule{0em}{1pt}{1pt}
     & EER  & $C_{llr}^\prime $ & $C_{llr}^{min}$ & EER & $C_{llr}^\prime$ & $C_{llr}^{min}$  \\
     \specialrule{0em}{1pt}{1pt}
      \hline 
    Clean & 1.65\%     &0.081& 0.060 & 1.63\% & 0.0.83& 0.063 \\ 
    
    Far-field & 3.84\%     & 0.156& 0.141 & 3.68\% & 0.151& 0.139 \\
    LENA-booth & 3.27\%     & 0.123& 0.107 & 3.23\% & 0.127& 0.109 \\
    LENA-field & 7.50\%     & 0.280 & 0.254 & 7.42\% & 0.289& 0.256 \\

    \hline
    \hline 
    \end{tabular}%
  
  \vspace{-1ex}
\end{table}%

Based on this motivation, we consider discrepancy minimization within an adaptation procedure at both the segment-level and frame-level to avoid inaccurate discrepancy estimation caused by the domain-wise embedding distribution deviation. Results from Table $\rm\uppercase\expandafter{\romannumeral6}$ show improvement with adaptation based on discrepancy minimization, where frame-level adaptation contributes to improvement with a slight EER reduction. Additionally, we utilize a zoo-plot visualization \cite{alexander2017zooplots} to explore a sample analysis on individual speakers, or speaker groups (see in Fig. 7). The zoo-plot shows a scatter type visualization based on mean values of both target and non-target scores for each speaker label, and speakers who fall within the four quartiles are assigned to animal groups (worms, chameleons, doves, and phantoms) with each set showing different characteristics. The black ellipses show mean values of all target and non-target scores for each domain, with the domain-index color in the center of each eclipse. Speakers toward the upper right corner have lower genuine variability and higher imposter variability. For example, speakers in the CRSS-Forensic LENA-field (recorded in 7 indoor and outdoor locations) tend to be more difficult to verify correctly than those in other domains. This visualization helps reveal potential algorithmic weaknesses against certain classes of speakers and domains. In terms of a statistics comparison, discrepancy minimization does improve EER of the pre-trained system with a relative decrease of 49.54\%, 19.83\%, 24.67\%, 23.09\% in each set, respectively. The average relative EER decrease is 29.28\%. 
\subsection{Reducing the Noise Mismatch Impact}
\begin{figure*}[ht]
  \centering
  \vspace{-3.5ex}
  \subfigure[discrepancy-minimization]{\includegraphics[width=0.8\linewidth]{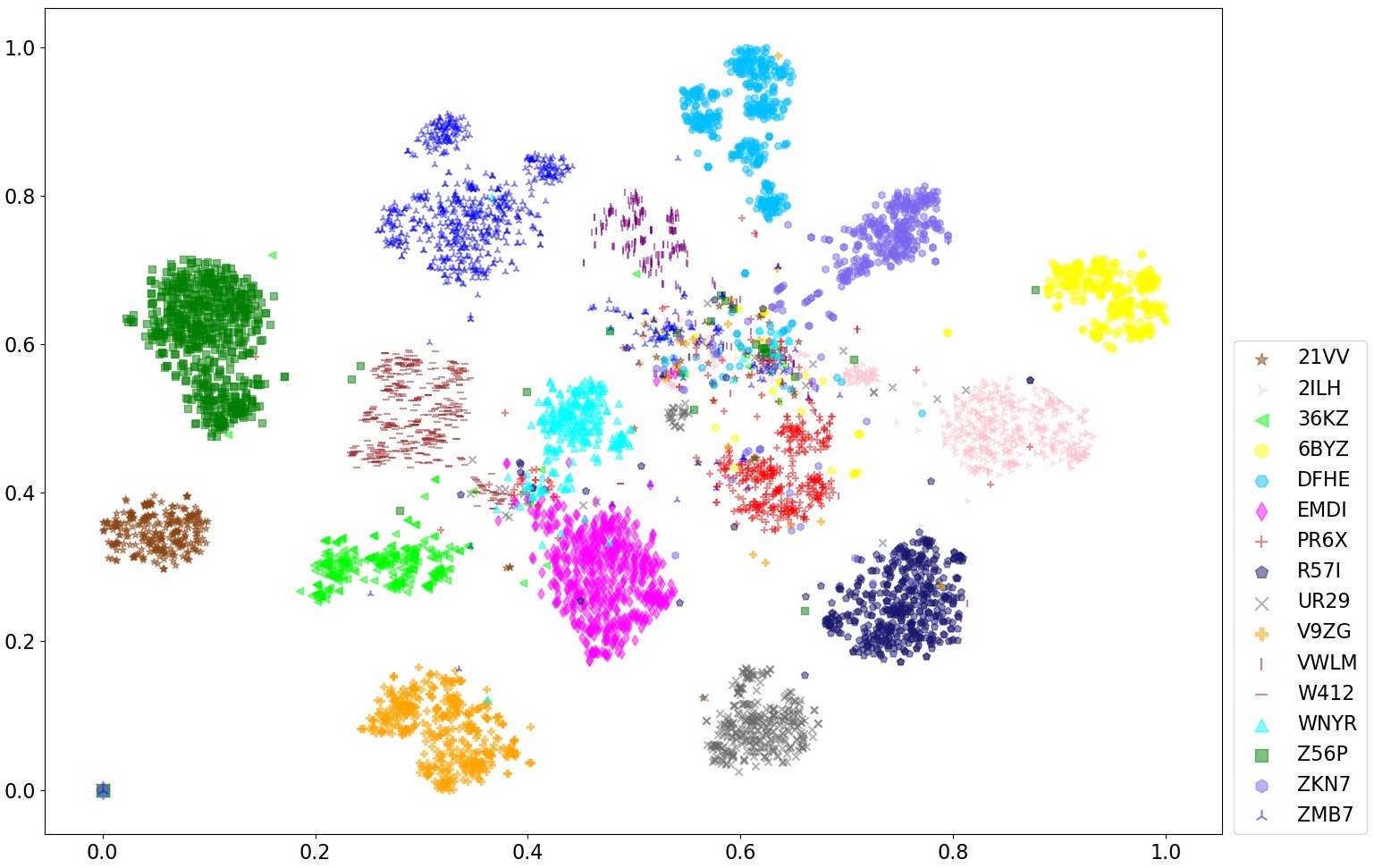}}

  \subfigure[moment-matching]{\includegraphics[width=0.8\linewidth]{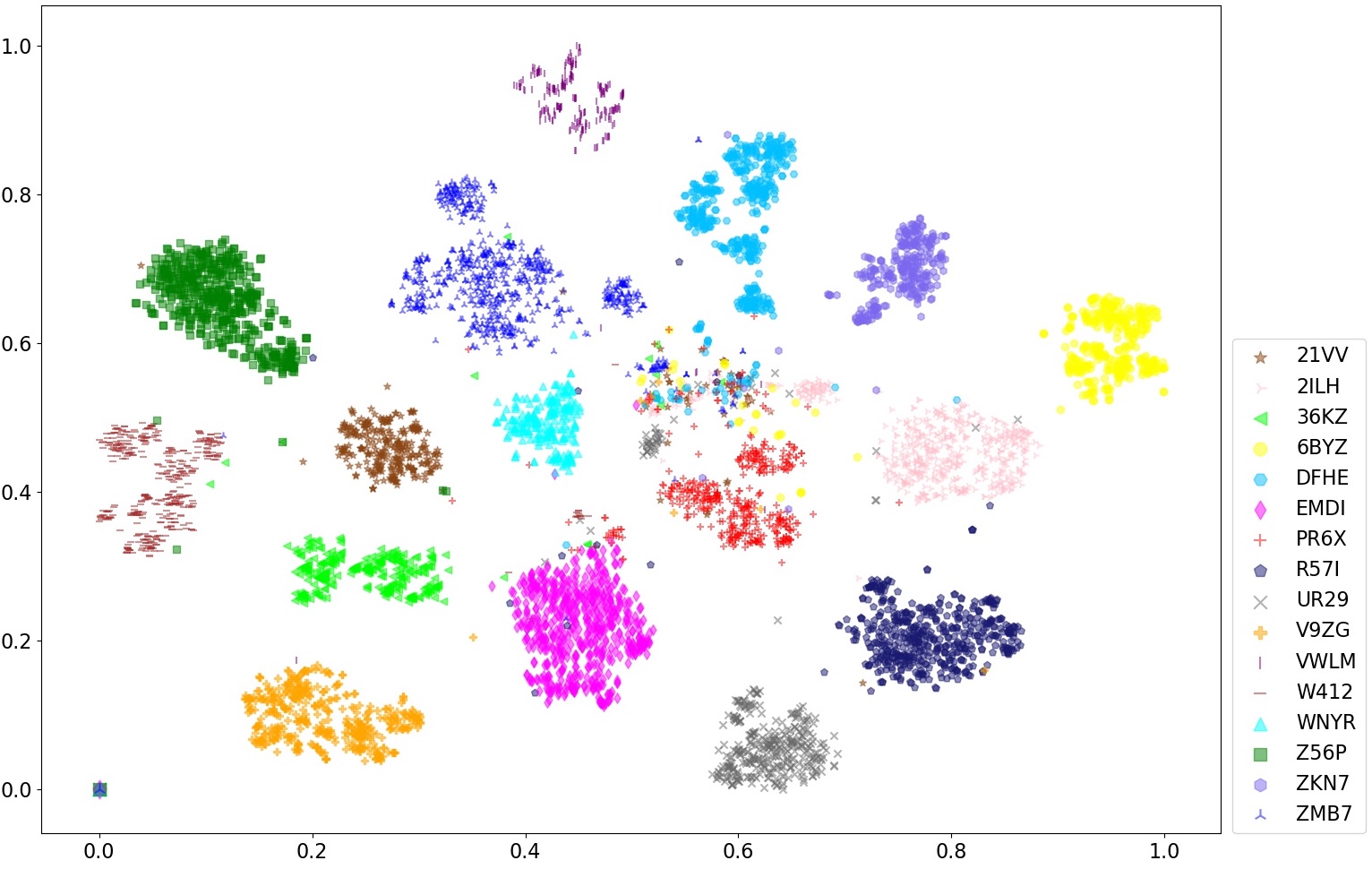}}
  \captionsetup{font={footnotesize}}
  \caption{Visualization of LENA-field speaker embeddings by discrepancy-minimizing (a) and moment-matching (b) systems using t-SNE.}
  \vspace{-3ex}
  
\end{figure*}
For the CRSS-Forensic corpus, the LENA-field portion consists of 7 diverse individual locations, which differ from the other three forensic portions obtained in the sound booth. As shown in the experiment results from Table $\rm\uppercase\expandafter{\romannumeral4}$, the noise mismatch exerts a significant impact on speaker recognition performance. Therefore, here we employ the moment-matching method with an adversarial training schema for adaptation to minimize both the domain shift and simultaneously mitigate the impact of noise. 

\begin{table}[ht]
  \centering
  \renewcommand{\arraystretch}{1.3}
  \captionsetup{font={footnotesize}}
  \caption{SR Result for Moment-matching System}
  \vspace{-1ex}
    \begin{tabular}{ccccc}
    \hline
    \hline 
     
     & EER & $C_{llr}$ & $C_{llr}^\prime$ & $C_{llr}^{min}$  \\
     \specialrule{0em}{1pt}{1pt}
      \hline 
    Clean & 1.50\%  & 0.173  & 0.08 & 0.060 \\ 
    
    Far-field & 3.45\%  & 0.276   & 0.14 & 0.131  \\
    LENA-booth & 3.05\%  & 0.211   & 0.111 & 0.100 \\
    LENA-field & 6.53\%  & 0.479   & 0.254 & 0.233 \\

    \hline
    \hline 
    \end{tabular}%
  \vspace{-1ex}
\end{table}%

The moment-matching advancement is shown to reduce EER of the pre-trained system with a relative decrease of 53.56\%, 24.84\%, 33.71\%, 27.38\% in each set, respectively. The average relative EER decrease is 34.87\%. As shown in Table $\rm\uppercase\expandafter{\romannumeral7}$, moment-matching with adversarial training improves SV performance for both sound-booth and LENA-field datasets by dynamically aligning the distribution of the LENA-field set with sound-booth captured audio sets. This distribution alignment is a mutual recalibration process, which suggests sound-booth data provides more distinguishable speaker information versus LENA-field data. Here, LENA-field data increases data diversity of sound-booth data for better generalization. Specifically, Table $\rm\uppercase\expandafter{\romannumeral8}$ summarizes the statistics for system SV results for each location of the LENA-field set. A comparison of Table $\rm\uppercase\expandafter{\romannumeral8}$ versus baseline x-Vector result (as shown in Table $\rm\uppercase\expandafter{\romannumeral4}$) confirms the dramatic benefits of the proposed solution. 
\begin{table}[h]
  \centering
  \renewcommand{\arraystretch}{1.3}
  \captionsetup{font={footnotesize}}
  \caption{SV Result for the Moment-matching System in 7 Locations}
  \vspace{-1ex}
    \begin{tabular}{ccccc}
    \hline
    \hline 
     7IOL&EER&$C_{llr}$&$C_{llr}^\prime$&$C_{llr}^{min}$\\
     \specialrule{0em}{1pt}{1pt}
     \hline
     Cafeteria&9.48\%&0.693&0.334&0.301\\
     Game room&9.36\%&0.702&0.358&0.324\\
     Hallway&6.13\%&0.491&0.248&0.221\\
     Lobby&5.17\%&0.377&0.212&0.181\\
     Office&3.30\%&0.254&0.132&0.118\\
     Parking Lot&4.19\%&0.279&0.177&0.152\\
     Walking Path&6.18\%&0.451&0.231&0.206\\
     
    \hline
    \hline 
    \end{tabular}%
   \vspace{-1ex}

\end{table}%

To visualize the effect of the moment-matching system on speaker recognition performance for LENA-field data, we further assess the quality of the learned speaker features using a t-distributed Stochastic Neighbor Embedding (t-SNE) plot \cite{maaten2008visualizing} (see in Fig. 8). We plot embeddings after LDA from 16 speakers of the CRSS-Forensic test set, which are generated by the discrepancy-minimizing system and moment-matching system. In the center of Fig. 8 (a), there is a cluster of outlier samples for the LENA-field speaker embeddings from the discrepancy-minimizing system, which confirms that they are easily misclassified with ambiguous identities. Alternatively, Fig. 8 (b) shows a sparse confusion cluster in the center which highlights how utilizing the domain-specific decision boundaries (noted in Sec. $\rm\uppercase\expandafter{\romannumeral6}.B$) works to improve speaker discrimination of outlier samples near the classification boundaries by dynamically aligning distributions in the moment-matching system. Several previous outlier samples in Fig. 8 (a) are also reclustered into corresponding groups in Fig. 8 (b), where most of the remaining samples could be actual outliers such as noise and speech of non-target speakers. Speech in the LENA-field will often contain sporadic noise and non-related speech due to the naturalistic field locations, which are also labeled as target speakers with a coarse-grained transcription. 

\section{Forensic Speaker Recognition}
The object of forensic speaker recognition is to assist in the "trier of fact" (i.e., a judge, a panel of judges, or a jury) in order to render a decision about the origin of a speech voice recording whose identity is in question. Systems with lower EER or LLR suggest that they are more capable to generate instructive scores with higher validity and reliability. We explored several speaker adaptation methods and compared their speaker recognition performance, which aim to achieve a reliable system able to produce a measure of evidence in the form of a likelihood ratio (LR) score as the strength of evidence. The LRs expresses the likelihood of the speech evidence under the two competing hypotheses (i.e. the prosecution hypothesis $H_0$: the suspected speaker is the same as the source of the questioned recording versus the defense hypothesis $H_1$: the suspected speaker is different from the source of the questioned recording \cite{drygajlo2012automatic}). The LRs is the ratio between these two statements $H_0$ and $H_1$.  

 The x-Vector modeling approach with a PLDA backend can be applied for calculating LRs. The goal of PLDA is to project data samples from the feature space to a latent space such that samples from the same class are modeled using the same distribution\cite{ioffe2006probabilistic}. Given $n$ utterance-level speaker embeddings $\lbrace\vct{u}_i^p\rbrace_{i=1}^n$ of speaker $p$ in the latent space and one utterance-level speaker embedding $\vct{u}^q$ of speaker $q$ in the latent space, if we need to find whether they belong to same speaker or not, then we compute the likelihood ratio R based on two hypothesis $H_0$ and $H_1$,
\begin{equation}
R(\lbrace\vct{u}_i^p\rbrace_{i=1}^n,\vct{u}^q)=\frac{likelihood(H_0)}{likelihood(H_1)}=\frac{P(\lbrace\vct{u}_i^p\rbrace_{i=1}^n,\vct{u}^q)}{P(\lbrace\vct{u}_i^p\rbrace_{i=1}^n)P(\vct{u}^q)},\label{eq:eq17}
\end{equation}
where
\begin{equation}
P(\lbrace\vct{u}_i^p\rbrace_{i=1}^n)=\int P(\vct{u}_1^p|\vct{v})...P(\vct{u}_n^p|\vct{v})P(\vct{v})d\vct{v}
\end{equation}
is the distribution of a set of examples,given that they belong to the same class, and $\vct{v}$ represents the class centers in the latent space. The log likelihood ratio $log(R)$ is known as PLDA scores. With larger $log(R)$ values, there is stronger support for the $H_0$ hypothesis, and with smaller $log(R)$ values, there is stronger support for the $H_1$ hypothesis. Here, we took 6 utterances for each speaker as enrollment data (here, $n=6$ in Eq. \eqref{eq:eq17}). Table $\rm\uppercase\expandafter{\romannumeral9}$ gives an example of quantitative measurement in the form of PLDA score generated by the Moment-Matching system between the given speakers and a specific recording. 
\begin{table}[h]
  \centering
  \renewcommand{\arraystretch}{1.3}
  \captionsetup{font={footnotesize}}
  \caption{PLDA scores between each speaker and a recording}
  \vspace{-1ex}
    \begin{tabular}{cccc}
    \hline
    \hline 
     &$speaker\_id$&$record\_id$&$PLDA\_score$\\
     
     \hline
     &21VV &$21VV\_LENA\_field\_100$ &7.585\\
	 &2ILH &$21VV\_LENA\_field\_100$ &-15.393\\
	 &36KZ &$21VV\_LENA\_field\_100$ &0.145\\
 	 &6BYZ &$21VV\_LENA\_field\_100$ &-13.596\\
     &DFHE &$21VV\_LENA\_field\_100$ &-15.123 \\
     &EMDI &$21VV\_LENA\_field\_100$ &-4.790\\
	 &PR6X &$21VV\_LENA\_field\_100$ &-11.299 \\
     &R57I &$21VV\_LENA\_field\_100$ &-9.353 \\
     &UR29 &$21VV\_LENA\_field\_100$ &-10.282\\
     &V9ZG &$21VV\_LENA\_field\_100$ &-0.368 \\
     &VWLM &$21VV\_LENA\_field\_100$ &-1.464 \\
     &W412 &$21VV\_LENA\_field\_100$ &-5.462 \\
     &WNYR &$21VV\_LENA\_field\_100$ &-5.688 \\
     &Z56P &$21VV\_LENA\_field\_100$ &-4.420\\
     &ZKN7 &$21VV\_LENA\_field\_100$ &-11.657\\
     &ZMB7 &$21VV\_LENA\_field\_100$ &-5.653 \\

    \hline
    \hline 
    \end{tabular}%
   \vspace{-1ex}

\end{table}%

Speech data from these 16 speakers constitute the entire test set. The suspected speakers during LRs calculation are called the relevant population in forensic speaker recognition. To avoid potential bias in the case proper, those speakers are often selected by a panel of listeners (e.g., police officers with linguistic background have no prior knowledge of a particular case)\cite{morrison2012database}. Since we already have the LRs, it can be interpreted based on the odds form of Bayes' theorem, which is represented as,
\begin{equation}
\frac{P(H_0|E)}{P(H_1|E)}=\frac{P(E|H_0)}{P(E|H_1)}\times\frac{P(H_0)}{P(H_1)},
\end{equation}
where $E$ represents the observed speech evidence. This Bayes' theorem shows how the LRs can be combined with prior knowledge concerning the case (knowledge unrelated to speech data) in order to arrive at posterior odds. Only the LRs is the province provided by the speaker recognition system; the prior odds and posterior odds are the province of the court. The forensic experts should only produce the LRs in actual court cases and leave prior odds to the court or jury to interpret or assess. The judge or the jury in the court can use such an non-categorical opinion for their deliberations and decision.

\maketitle

\section{Conclusion}

For forensic speaker recognition, addressing mismatch due to naturalistic field locations is a significant challenge. In general, fine-tuning is commonly employed for network model adaptation when a domain mismatch exists between train and test data. However, that approach usually considers only a single domain mismatch. In practical scenarios for forensic audio analysis, speech data are typically collected in multiple acoustic environments, which offer unique challenges to speaker recognition system development due to location uncertainty and diverse mismatch between reference and naturalistic field recordings. A speaker recognition system can deliver different speaker discrimination performance while evaluated on the dataset collected from multiple acoustic environments. In this study, we adopted a domain adversarial training (DAT) method with a gradient reversal layer to learn domain-invariant and speaker-discriminative representations. The DAT gives competitive results on each domain. Additionally, we formulated a discrepancy-minimizing solution to perform model adaptation for the purpose of improving speaker recognition performance across each potential field location with an overall smaller domain discrepancy. As demonstrated in our results, the solution improves speaker recognition system performance for each domain, which demonstrates that minimizing the domain discrepancy at both the frame-level and segment-level benefits system speaker discrimination. However, this improvement can still be unbalanced, as was shown with a higher EER result for the LENA-field set versus others. The LENA-field data was collected in locations entirely different from environments of the other three datasets. Accordingly, we proposed a moment-matching solution with an adversarial training schema for model adaptation to minimize domain discrepancy and simultaneously mitigate the impact of noise for LENA-field data with the help of sound-booth captured audio. Consequently, the moment-matching system achieved the best speaker recognition results for each domain, with absolute EERs of 1.50\%, 3.45\%, 6.53\%, 3.05\% for the Clean, Far-field, LENA-field, and LENA-booth sets, respectively. Overall, the learned speaker representations through domain adversarial training (DAT), discrepancy-minimizing, and moment-matching solutions are less dependent on shifts in acoustic domains, which provides a solution to the challenging multi-source domain adaptation problem in forensic speaker recognition. Finally, we applied the most effective overall system for an independent simulative forensic case to show how the system solution can support the judge or jury in a court scenario to make a decision with a strength-of-evidence statement in the form of a likelihood ratio.


%




%
\bibliographystyle{IEEEtran}
\bibliography{IEEEexample.bib}

%

\begin{IEEEbiography}
[{\includegraphics[width=1in,height=1.25in,clip,keepaspectratio]{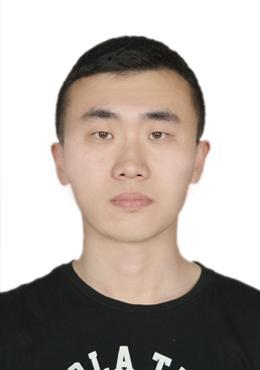}}]{Zhenyu Wang}
received the B.S. degree majoring in digital media technology from Hangzhou Dianzi University, Hangzhou, China in 2015. He received the M.S. degree in engineering computer system architecture from Beijing Language and Culture University, Beijing, China in 2019. He started working troward the Ph.D. degree in computer engineering at the University of Texas at Dallas (UTD), Richardson, TX, USA, in 2019. He works with Professor John H.L. Hansen at the Center for Robust Speech Systems.  Since the same year, he has been a Graduate Research Assistant with the Center for Robust Speech Systems (CRSS), UTD. His research interests include mispronunciation verification, computer-assisted language learning, forensic audio analysis and model adaptation for open-set speaker recognition system, representation learning used for acoustic modeling. He has authored around five journal and conference papers in the field of speech processing and language technology. 
\end{IEEEbiography}

\begin{IEEEbiography}[{\includegraphics[width=1in,height=1.25in,clip,keepaspectratio]{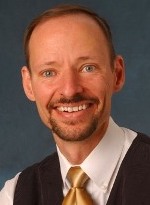}}]{John H. L. Hansen}
John H. L. Hansen (S’81–M’82–SM’93–F’07) received the B.S.E.E. degree from the College of Engineering, Rutgers University, New Brunswick, NJ, USA, in 1982, the M.S. and Ph.D. degrees in electrical engineering from Georgia Institute of Technology, Atlanta, GA, USA, in 1983 and 1988, respectively. He joined Erik Jonsson School of Engineering and Computer Science, University of Texas at Dallas (UTDallas), in 2005, where he currently serves as a Jonsson School Associate Dean for Research, as well as a Professor of Electrical and Computer Engineering, the Distinguished University Chair in Telecommunications Engineering, and a joint appointment as a Professor in the School of Behavioral and Brain Sciences (Speech \& Hearing). He previously served as a UTDallas Department Head of Electrical Engineering from August 2005 to December 2012, overseeing a +4× increase in research expenditures (4.5 to 22.3 M) with a 20\% increase in enrollment along with hiring 18 additional T/TT faculty, growing UTDallas to the eighth largest EE program from ASEE rankings in terms of degrees awarded. At UTDallas, he established the Center for Robust Speech Systems (CRSS). Previously, he served as a Department Chairman and Professor of Speech, Language and Hearing Sciences, and a Professor in Electrical and Computer Engineering, University of Colorado-Boulder (1998–2005), where he co-founded and served as an Associate Director of the Center for Spoken Language Research. In 1988, he established the Robust Speech Processing Laboratory and continued to direct research activities in CRSS at UTDallas. He is the author/coauthor of 800+ journal and conference papers including 13 textbooks in the field of speech processing and language technology, signal processing for vehicle systems, coauthor of textbook: Discrete-Time Processing of Speech Signals (IEEE Press, 2000), co-editor of DSP for In-Vehicle and Mobile Systems (Springer, 2004), Advances for In-Vehicle and Mobile Systems: Challenges for International Standards (Springer, 2006), In-Vehicle Corpus and Signal Processing for Driver Behavior (Springer, 2008), and lead author of the report The Impact of Speech Under Stress on Military Speech Technology, (NATO RTOTR-10, 2000). He has supervised 95 Ph.D./M.S. thesis candidates (54 Ph.D., 41 M.S./M.A.). His research interests include the areas of digital speech processing, analysis and modeling of speech and speaker traits, speech enhancement, feature estimation in noise, signal processing for hearing impaired/cochlear implants, robust speech recognition with emphasis on machine learning and knowledge extraction, and in-vehicle interactive systems for hands-free human-computer interaction. Dr. Hansen received the honorary degree Doctor Technices Honoris Causa from Aalborg University (Aalborg, DK) in April 2016, in recognition of his contributions to speech signal processing and speech/language/hearing sciences. He was recognized as an IEEE Fellow (2007) for contributions in “Robust Speech Recognition in Stress and Noise,” International Speech Communication Association (ISCA) Fellow (2010) for contributions on research for speech processing of signals under adverse conditions, and received The Acoustical Society of Americas 25 Year Award (2010) in recognition of his service, contributions, and membership to the Acoustical Society of America. He previously served as ISCA President (2017–2021) and Vice-president (2015-2017) and currently serves as tenure and member of the ISCA Board, having previously served as the Vice-President (2015–2017). He also is serving as a Vice-Chair on U.S. Office of Scientific Advisory Committees (OSAC) for OSAC-Speaker in the voice forensics domain (2015–2017). Previously, he served as an IEEE Technical Committee (TC) Chair and member of the IEEE Signal Processing Society: Speech-Language Processing Technical Committee (SLTC) (2005–2008; 2010–2014; elected IEEE SLTC Chairman for 2011–2013, Past-Chair for 2014), and elected as an ISCA Distinguished Lecturer (2011–2012). He has served as the member of the IEEE Signal Processing Society Educational Technical Committee (2005–2008; 2008–2010); Technical Advisor to the U.S. Delegate for NATO (IST/TG-01); IEEE Signal Processing Society Distinguished Lecturer (2005–2006), Associate Editor for IEEE TRANSACTION SPEECH AND AUDIO PROCESSING (1992–1999), Associate Editor for IEEE SIGNAL PROCESSING LETTERS (1998–2000), Editorial Board Member for IEEE SIGNAL PROCESSING MAGAZINE (2001–2003); and Guest Editor (October 1994) for special issue on Robust Speech Recognition for IEEE TRANSACTION SPEECH AND AUDIO PROCESSING. He is serving as an Associate Editor for JASA, and served on Speech Communications Technical Committee for Acoustical Society of America (2000–2003). He was the recipient of The 2005 University of Colorado Teacher Recognition Award as voted on by the student body. He organized and served as the General Chair for ISCA INTERSPEECH-2002, September 16–20, 2002, Co-Organizer, and Technical Program Chair for IEEE ICASSP-2010, Dallas, TX, USA, March 15–19, 2010, and Co-Chair and Organizer for IEEE SLT-2014, December 7–10, 2014 in Lake Tahoe, NV, USA.
\end{IEEEbiography}





\end{document}